\documentclass[a4paper, 11pt]{article}\usepackage[]{graphicx}\usepackage[table]{xcolor}
\makeatletter
\def\maxwidth{ %
  \ifdim\Gin@nat@width>\linewidth
    \linewidth
  \else
    \Gin@nat@width
  \fi
}
\makeatother

\definecolor{fgcolor}{rgb}{0.345, 0.345, 0.345}

\usepackage{framed}
\makeatletter
 {\par\unskip\endMakeFramed%
 \at@end@of@kframe}
\makeatother

\definecolor{shadecolor}{rgb}{.97, .97, .97}
\definecolor{messagecolor}{rgb}{0, 0, 0}
\definecolor{warningcolor}{rgb}{1, 0, 1}
\definecolor{errorcolor}{rgb}{1, 0, 0}
\newenvironment{knitrout}{}{} 

\usepackage{alltt}

\usepackage[scaled]{helvet}

\usepackage{geometry}
\usepackage{hyperref}
\hypersetup{
  bookmarksopen=true,
  breaklinks=true,
  colorlinks=true,
  linkcolor=black,
  anchorcolor=black,
  citecolor=blue,
  urlcolor=blue,
}

\usepackage{amsmath, amsfonts, amsthm, amssymb}
\usepackage{natbib}
\usepackage{pdflscape}
\usepackage{csquotes}
\usepackage{tabularx}
\usepackage{booktabs}
\usepackage[figuresright]{rotating}
\usepackage{afterpage}

\usepackage{authblk}  
\usepackage{lmodern}  
\usepackage{setspace} 
\usepackage[T1]{fontenc}
\usepackage[utf8]{inputenc}
\usepackage[english]{babel}

\usepackage{algorithm}
\usepackage{algpseudocode}
\algtext*{EndProcedure}

\usepackage[table]{xcolor}
\usepackage{graphicx}
\usepackage{multirow}

\newcommand{\vertlabel}[2]{%
  \rotatebox[origin=c]{90}{%
    \colorbox{gray!20}{%
      \parbox[c][1.2cm][c]{#1}{\centering\bfseries #2}%
    }%
  }%
}

\usepackage{caption}
\newcommand{\hti}{{\widehat{\theta}_i}} 
\newcommand{\ti}{{\theta_i}} 
\newcommand{\tn}{{\theta_{\scriptscriptstyle{\text{new}}}}} 
\newcommand{\tnj}{{\theta_{\scriptscriptstyle{\text{new}}}^{(j)}}} 
\newcommand{\hmu}{{\widehat{\mu}}} 
\newcommand{\hmuivw}{{\widehat{\mu}_{\scriptscriptstyle{\text{IVW}}}}} 
\newcommand{\hsi}{{\widehat{\sigma}^2_i}} 
\newcommand{\taus}{{\tau^2}} 
\newcommand{\htaus}{{\widehat{\tau}^2}} 
\newcommand{\wit}{{w_i(\taus)}}
\renewcommand{\d}{{\partial}}

\DeclareMathOperator{\se}{se}  
\DeclareMathOperator{\Var}{Var} 
\DeclareMathOperator{\Q}{Q}
\DeclareMathOperator{\E}{\mathsf{E}}
\DeclareMathOperator{\Nor}{N} 
\DeclareMathOperator{\Chi}{\chi^2}

\DeclareMathOperator{\SN}{SN}
\DeclareMathOperator{\Uni}{U}

\renewcommand{\d}{{\mathrm{d}}}

\newcommand{\taustar}{{{\tau}_{(b)}^{2*}}}
\newcommand{\mustar}{{{\mu}_{(b)}^{*}}}
\newcommand{\tnstar}{{{\theta}_{\scriptscriptstyle{\text{new}}}^{*(b)}}}
\newcommand{\tnstarone}{{{\theta}_{\scriptscriptstyle{\text{new}}}^{*(1)}}}
\newcommand{\tnstartwo}{{{\theta}_{\scriptscriptstyle{\text{new}}}^{*(2)}}}
\newcommand{\tnstarB}{{{\theta}_{\scriptscriptstyle{\text{new}}}^{*(1)}, \ldots, {\theta}_{\scriptscriptstyle{\text{new}}}^{*(B)}}}
\newcommand{\tnstarplus}{{{\theta}_{\scriptscriptstyle{\text{new}}}^{*(b  + 1)}}}

\newcommand{\coverage}{{\text{Cov}}}
\newcommand{\semc}{{\se_{\scriptscriptstyle{\text{MC}}}}}

\newtheorem{definition}{Definition}

\IfFileExists{upquote.sty}{\usepackage{upquote}}{}
\begin{document}
\title{\bfseries Prediction intervals for random-effects meta-analysis based on confidence distributions and Edgington's method\\[1em]}

\author{
David Kronthaler$^{1}$\thanks{david.kronthaler@uzh.ch} \quad and \quad Leonhard Held$^{1}$\thanks{leonhard.held@uzh.ch}\\

$^{1}$Epidemiology, Biostatistics and Prevention Institute,
Department of Biostatistics, University of Zurich,
Hirschengraben 84, 8001 Zurich, Switzerland
}

\date{}

\maketitle

\begin{abstract}
Statistical inference about the average effect in random-effects meta-analysis has been considered insufficient in the presence of substantial between-study heterogeneity. Predictive distributions are well-suited for quantifying heterogeneity since they are interpretable on the effect scale and provide clinically relevant information about future events. We construct predictive distributions accounting for uncertainty through confidence distributions from Edgington's $p$-value combination method and the generalized heterogeneity statistic. Simulation results suggest that 95\% prediction intervals typically achieve nominal coverage when more than three studies are available and effectively reflect skewness of effect estimates in scenarios with 20 or less studies. Formulations that ignore uncertainty in heterogeneity estimation typically fail to achieve correct coverage, underscoring the need for this adjustment in random-effects meta-analysis.\\[0.5em]

\end{abstract}

\newpage
\section{Introduction}

Meta-analysis describes statistical methodology for synthesizing evidence across independent studies. Under a random-effects model, variation in study estimates is attributed to sampling variability and variability of true effects across studies, with the commonly used estimated mean effect reflecting the average of the distribution of true effects. However, inference about the average effect only 
does not help to understand the variability of effects across studies 
\citep{Riley2011, IntHout2016}. \citet[][p.38]{Viechtbauer2007} emphasizes that \enquote{quantifying the amount of heterogeneity and exploring its sources are among the most important aspects of systematic reviews}.
Traditional heterogeneity measures such as the estimated between-study variance and Higgins' $I^2$ have been criticized for their limited interpretability and sensitivity to study sample sizes. For instance, Higgins' $I^2$, being a relative measure, does not convey the absolute effect variation, which is typically of interest in practice \citep{borenstein2017basics}.

Predictive distributions of future effects are considered more useful for quantifying heterogeneity as they express variability on the effect measure scale and inform about clinically relevant future events \citep{Higgins2008, IntHout2016}. 
Frequentist predictive distributions are commonly summarized using prediction intervals, and several authors have advocated for reporting these in random-effects meta-analyses \citep{Riley2011, Guddat2012, IntHout2016}. \citet[][p.141]{Higgins2008} suggest that \enquote{predictive distributions are potentially the most relevant and complete statistical inferences to be drawn from a random-effects meta-analysis}. Prediction intervals frequently suggest strong variability in treatment responses. According to \citet{IntHout2016}, in approximately 70\% of random-effects meta-analyses in the Cochrane database with 95\% confidence intervals not covering a zero-effect, 95\% prediction intervals indicate that future effects may be zero or opposite in direction to the estimated average effect; in 20\%, they include effects that are opposite in direction and at least as large in absolute magnitude as the estimated average effect.

Despite growing recognition of the need for meta-analytic methods that account for skewness of study estimates \citep{higgins2008meta, noma2022meta}, most existing methods for constructing frequentist predictive distributions constrain them to be symmetric around the estimated average effect. 
These methods tend to be poorly calibrated for skewed effect distributions, with prediction intervals often not achieving nominal coverage \citep{matrai_archive}. Further, uncertainty about the heterogeneity estimate is typically addressed by applying a $t$-distribution, but this \emph{post hoc} adjustment is relatively arbitrary, especially concerning the degrees of freedom, with several choices proposed \citep{Partlett2016, Veroniki2018}. Simulation studies indicate that predictive distributions adjusted in that manner may be poorly calibrated in meta-analyses with few studies or considerable heterogeneity \citep{Partlett2016, Wang2019}.

To address these limitations, we propose constructing predictive distributions using Edgington's method based on the sum of $p$-values \citep{edgington1972additive}. Classical meta-analysis employs a weighted average of study estimates, but can be generalized as combining $p$-value functions or, equivalently, confidence distributions across studies \citep{xie2011confidence}, with the weighted Stouffer method \citep{Senn2021} recovering the classical estimator. \citet{heldpawelhofman2024} suggested that alternative $p$-value combination methods may serve as substitutes or complements to the classical approach. In particular, Edgington's method has been recommended because it is invariant to the orientation of the alternative under which one-sided $p$-values are constructed and yields nearly unbiased estimates. Importantly, it produces close-to-nominal coverage confidence intervals that adapt to skewness of effect estimates, preserving asymmetry rather than imposing symmetric summaries.

Our approach accounts for uncertainty about the average effect estimate through a confidence distribution constructed with Edgington's method, and similarly accounts for uncertainty about the heterogeneity estimate through its confidence distribution implied by the generalized heterogeneity statistic \citep{Viechtbauer2007}. We illustrate the methodology with a case study and evaluate their performance in a simulation study.

\section{Methods}

\subsection{The Random-Effects Model}

The proposed predictive distributions are constructed under a random-effects model assuming that future effects $\tn$ and true effects of $k$ observed studies $\ti, i \in \{1, \dots, k\}$, are exchangeable and jointly normally distributed around mean $\mu$ with variance $\taus$. The effect estimates $\hti$ are assumed to be normally distributed around $\ti$ with variance $\hsi$, the squared standard error from study $i$, which is treated as fixed. This hierarchical $k + 2$ parameter model collapses to a two-parameter model by marginalization:
\begin{equation}\label{eq:rema}
\ti  \sim \Nor(\mu, \taus), \ \hti\mid\ti \sim \Nor(\ti, \hsi) \ \Rightarrow \ \hti \sim \Nor(\mu, \taus + \hsi).
\end{equation}
Under the exchangeability assumption, the target of prediction coincides with the estimation of the distribution of true effects. In contrast, predictive distributions should not be used to quantify heterogeneity under a fixed-effects model \citep{rice2018re}, because in this framework $\ti$ and $\tn$ are not connected via a distribution.

\subsection{Prior Work on Predictive Distributions}

To our knowledge, the earliest predictive distribution was proposed by \citet{Skipka2006}, and can be understood as integrating the normal effect distribution over the normal sampling distribution of the inverse-variance weights (IVW) estimator $\hmuivw$ of the average effect,
\[
\int_\mathbb{R} f_{\Nor}(\tn \mid \mu', \htaus) \ f_{\Nor}(\mu' \mid \hmuivw, \se(\hmuivw)^2) \ \d \mu',
\]
where $f_{\Nor}$ denotes a normal density, yielding the predictive distribution $\Nor(\hmuivw, \htaus + \se(\hmuivw)^2)$. To account for uncertainty in $\htaus$, the Higgins--Thompson--Spiegelhalter (HTS) method applies a $t$-distribution with $k-2$ degrees of freedom \citep{Higgins2008}:
\[
\tn \sim \hmuivw + t_{k-2} \sqrt{\htaus + \se(\hmuivw)^2}.
\]
The degrees of freedom $k - 2$ reflect uncertainty in two parameter estimates, but several alternatives using different degrees of freedom or standard errors have been proposed \citep{Veroniki2018}. Simulation studies suggest limitations in this heterogeneity uncertainty adjustment \citep{Partlett2016, matrai_archive}, and \citet{Nagashima2018} proposed a parametric bootstrap procedure based on the standard error ${\se_{\scriptscriptstyle{\text{HKSJ}}}}(\hmuivw)$ of the IVW estimator $\hmuivw$ by \citet{Hartung2001} and \citeauthor{sidik2002simple} (\citeyear{sidik2002simple}; HKSJ),
\[
\tn = \hmuivw + Z \cdot \tau - T \cdot {\se_{\scriptscriptstyle{\text{HKSJ}}}}(\hmuivw),
\]
where $Z \sim \Nor(0,1)$ and $T \sim t_{k-1}$. A sample $\tn$ is generated by drawing $\taus$ from its confidence distribution implied by the exact (generalized $\chi^2$) distribution of Cochran's $\Q$ statistic, recomputing $\hmuivw$ and ${\se_{\scriptscriptstyle{\text{HKSJ}}}}(\hmuivw)$, and sampling independent random variables $Z$ from a standard normal and $T$ from a $t_{k-1}$-distribution. Further, \citet{Wang2019} proposed a non-parametric approach based on the empirical distribution of ensemble estimates $\hti^*$ of $\ti$ by \citet{louis1991using}:
\[
\hti^* = \hmuivw + \sqrt{\frac{\htaus}{\htaus + \hsi}} \ (\hti - \hmuivw).
\]
The ensembles are centered at $\hmuivw$ and calibrated to have asymptotic variance $\taus$. Their empirical distribution is therefore an estimate of the predictive distribution. However, \citet{matrai_archive} reported poor coverage of prediction intervals derived from this approach in scenarios with few studies, potentially due to not adjusting for estimation uncertainty and relying too strongly on the asymptotic argument.

\subsection{\textit{P}-Value Functions and Confidence Distributions}

A $p$-value function treats the $p$-value as a function of the parameter of interest \citep{Fraser2019}. Consider the Wald test 
for a parameter $\mu$ with estimator $\hmu$ and standard normal pivot $Z(\mu) = (\hmu - \mu)/\se(\hmu)$. The corresponding one-sided (1s) and two-sided (2s) $p$-value functions are $p_{\text{1s},+}(\mu) = 1 - \Phi\left\{Z(\mu)\right\}$ for the alternative "greater", $p_{\text{1s},-}(\mu) = \Phi\left\{Z(\mu)\right\}$ for the alternative "less", and $p_{\text{2s}}(\mu) = 2 \min\left\{p_{\text{1s},+}(\mu),\; p_{\text{1s},-}(\mu)\right\}$, where $\Phi$ denotes the standard normal cumulative distribution function (CDF). $P$-value functions provide evidence against all possible null hypotheses (e.g., $p_{\text{1s},+}(\mu_0)$ equals the one-sided $p$-value for testing $H_0\colon\mu \le \mu_0$ vs.\ $H_1\colon\mu > \mu_0$) and can be used for both point and interval estimation. Importantly, the increasing one-sided $p$-value function $p_{\text{1s},+}(\mu)$ is typically the CDF of an (asymptotic) confidence distribution, with exceptions being e.g. strictly conservative one-sided tests (see Section~2.3.2 in \citealp{Xie2013}).

Confidence distributions are frequentist probability distributions over the parameter space, constructed without invoking prior distributions \citep{Cox1958}. Rather than representing the inherent distribution of a parameter, they are modernly interpreted as sample-dependent distributional summaries of uncertainty \citep{Xie2013}. One interpretation sees them as encompassing all possible confidence intervals simultaneously, where the \emph{confidence} or \emph{confidence probability} assigned to a parameter subspace corresponds to the confidence level of the interval spanning it \citep{Marschner2024}.
Formally, a confidence distribution is defined by \citet{SCHWEDER2002} as:
\begin{definition}[Confidence Distribution]\label{def:cd}
Let $\mathbf{Y}$ be a random vector with sample space $\mathcal{Y}$ and realization $\mathbf{y}$, and let $\mu \in \Theta$ be the parameter of interest. A function $C(\mathbf{Y}, \cdot)\colon\Theta \rightarrow [0, 1]$ is called a \emph{confidence distribution} of $\mu$ if (i) for each fixed $\mathbf{y} \in \mathcal{Y}$, $C(\mathbf{y}, \cdot)$ is a cumulative distribution function on $\Theta$, and (ii) at the true parameter value $\mu = \mu_0$, $C(\mathbf{Y}, \mu_0)$ follows a standard uniform distribution: $C(\mathbf{Y}, \mu_0) \sim \Uni(0,1)$. The function $C(\mathbf{Y}, \cdot)$ is called an \emph{asymptotic confidence distribution} if $C(\mathbf{Y}, \mu_0)$ converges in distribution to the standard uniform as the size of $\mathbf{Y}$ increases. The \emph{confidence density} is the derivative of the confidence distribution with respect to $\mu$: $c(\mathbf{Y}, \mu) = \partial_\mu C(\mathbf{Y}, \mu)$.
\end{definition}
An example illustrating $p$-value functions and confidence distributions for the Wald test is provided in Web Appendix~A.

\subsection{Meta-Analysis based on \textit{p}-Value Functions}
The marginal distributions under model~\eqref{eq:rema}, $\hti \sim \Nor(\mu, \taus + \hsi)$, induce the $k$ pivots and corresponding one-sided $p$-value functions for the alternative "greater",
\begin{equation}\label{eq:marginal}
Z_i(\mu) = \frac{\hti - \mu}{\sqrt{\taus + \hsi}} \sim \Nor(0, 1), \quad p_{\text{1s},+,i}(\mu) = 1 - \Phi\left\{Z_i(\mu)\right\},
\end{equation}
which are combined to yield a combined $p$-value function for $\mu$. One-sided $p$-value functions are preferred over two-sided $p$-value functions, since the latter may yield empty confidence intervals or confidence sets consisting of non-overlapping intervals \citep{heldpawelhofman2024}. Edgington's method of combining $p$-values corresponds to evaluating the CDF of the Irwin--Hall distribution (i.e., the distribution of the sum of $k$ independent standard uniform random variables) with parameter $k$ at the sum of $p$-values $s(\mu) = \sum_{i = 1}^k p_{\text{1s},+,i}(\mu)$. For $k \ge 12$, Edgington's method is approximated using a normal distribution based on a central limit theorem argument to mitigate overflow problems. Therefore, Edgington's combined $p$-value function is
\begin{equation}
p_E(\mu) = \begin{cases}
\frac{1}{k!} \sum_{j = 0}^{\lfloor s(\mu) \rfloor}(-1)^j \binom{k}{j} \{s(\mu) - j\}^k & \text{if } k < 12, \\
\Phi[\sqrt{12k} \ \{s(\mu)/k - 1/2\}] & \text{if } k \ge 12,
\end{cases}\label{eq:pEdgington}
\end{equation}
where the floor function $\lfloor s(\mu) \rfloor$ denotes the greatest integer less than or equal to $s(\mu)$. Edgington's combined $p$-value function provides a point estimator and a confidence intervals for $\mu$, analogous to the illustration for a single Wald test in Web Appendix~A.

To relate the $p$-value combination approach to classical meta-analysis, we note that, with appropriate weights, Stouffer's method recovers the  classical random-effects estimator as the solution to the combined $Z$-score equation based on $Z_i(\mu)$ from~\eqref{eq:marginal}. The weighted Stouffer statistic is $Z_S(\mu) = {\sum_{i=1}^k w_i Z_i(\mu)}/\sqrt{\sum_{i=1}^k w_i^2}$ \citep{Senn2021}. Choosing $w_i = 1/\sqrt{\taus + \hsi}$ and solving $Z_S(\mu) = 0$ for $\mu$ yields $\widehat{\mu} = \big(\sum_{i=1}^k \frac{\hti}{\taus + \hsi}\big)\big/\big(\sum_{i=1}^k \frac{1}{\taus + \hsi}\big)=:\hmuivw$.

\subsection{Predictive Distributions using Edgington's Method}

Assuming exchangeability of $\tn$ and $\ti$, $i \in \{1,\dots,k\}$, the basic form of a predictive distribution under the model in~\eqref{eq:rema} is a $\Nor(\mu, \taus)$. To account for estimation uncertainty of $\mu$ and $\taus$, we employ a confidence distribution approach. A confidence distribution of $\mu$ with confidence density $c(\mu \mid \taus)$ is given by Edgington's combined $p$-value function \citep{xie2011confidence}, which is conditional on the heterogeneity parameter $\tau^2$ through the study-specific pivots in \eqref{eq:marginal}. A simple predictive (confidence) distribution, \emph{PCD-fixed}, accounting for uncertainty in the estimation of $\mu$, is obtained by integrating the normal effect distribution over Edgington's confidence distribution using a plug-in estimate $\widehat{\tau}^2$ for the heterogeneity parameter:
\begin{equation}\label{eq:CDstandard}
c_{\text{fixed}}(\tn\mid \htaus) = \int_\mathbb{R} f_{\Nor}(\tn\mid\mu, \htaus) \ c(\mu \mid \htaus) \ \d\mu.
\end{equation}
This approach resembles those of standard predictive distributions, but differs in integrating over the confidence distribution of the parameter $\mu$ rather than a sampling distribution of the estimator $\hmuivw$. Marginalizing over a confidence distribution does not generally ensure that the resulting distribution retains valid frequentist properties, but is typically only an approximation, whose accuracy should be assessed via simulation \citep{Schweder2016}. Under the normal model with pivots for location and scale, which provides a setting closely related to the random-effects model in~\eqref{eq:rema}, integration of the joint confidence distribution can yield valid marginal confidence distributions \citep{pawitan2021}. Moreover, \citet{Nagashima2018} construct prediction intervals by sampling the heterogeneity parameter from its confidence distribution based on the exact distribution of Cochran's $\Q$, and demonstrate good performance of the resulting prediction intervals.

The predictive distribution in~\eqref{eq:CDstandard} relies on a fixed estimate $\htaus$ and ignores uncertainty in its estimation. We incorporate heterogeneity estimation uncertainty by integrating over a confidence distribution of $\taus$, implied by the generalized heterogeneity statistic 
\begin{equation}\label{eq:genQ}
\Q(\taus) = \sum_{i=1}^k w_i(\tau^2)
\big\{\widehat{\theta}_i - \widehat{\mu}_{\mathrm{IVW}}(\tau^2)\big\}^2,
\end{equation}
where $w_i(\taus) = (\taus + \hsi\big)^{-1}$ and 
$\widehat{\mu}_{\mathrm{IVW}}(\tau^2) = \big\{\sum_{i=1}^k w_i(\tau^2)\,\widehat{\theta}_i\big\}/\big\{\sum_{i=1}^k w_i(\taus)\big\}$ is the IVW random-effects estimator as a function of $\tau^2$ \citep{Viechtbauer2007}. Under the model in~\eqref{eq:rema}, $\Q(\taus)$ follows a $\chi_{k-1}^2$-distribution, and it is therefore a pivot in $\taus$ and yields a confidence distribution of $\taus$. This distribution is exact, but relies on known within-study variances and therefore only constitutes an approximate confidence distribution in practice.

Using~\eqref{eq:genQ} to construct confidence intervals for $\taus$ is referred to as the $\Q$-profile method \citep{Viechtbauer2007, Jackson2016}. Treating~\eqref{eq:genQ} as a confidence distribution of $\taus$ can be viewed as an extension, representing the set of all possible confidence intervals derived via the $\Q$-profile method. Note that this differs from the approach of \citet{Nagashima2018}, who use the exact confidence distribution of the standard $\Q$ statistic without extending it to the generalized version that varies with $\taus$.
We incorporate uncertainty in the estimation of heterogeneity into the predictive distributions by integrating over the confidence distribution implied by $\Q(\taus)$. The \emph{PCD-simplified} version
\begin{equation}\label{eq:PD_CD_TAU2_UV}
c_{\text{simplified}}(\tn) = \int_\mathbb{R^+} \int_\mathbb{R} f_{\Nor}(\tn \mid \mu, \taus) \ c(\mu \mid \htaus) \ c(\taus) \ \d\mu \ \d\taus
\end{equation}
uses a plug-in estimate $\htaus$ within Edgington's confidence distribution, whereas the \emph{PCD-full} version fully integrates over the conditional distribution of $\mu$:
\begin{equation}\label{eq:cdpdfull}
c_{\text{full}}(\tn) = \int_\mathbb{R^+} \int_\mathbb{R} f_{\Nor}(\tn \mid \mu, \taus) \ c(\mu \mid \taus) \ c(\taus) \ \d\mu \ \d\taus.
\end{equation}

Deterministic evaluation of the integrals in~\eqref{eq:PD_CD_TAU2_UV} and~\eqref{eq:cdpdfull} is computationally demanding. We propose a Monte Carlo algorithm to generate samples of the three unknown quantities $\taus$, $\mu$ and $\tn$. For each draw $b \in \{1,\ldots,B\}$, let $\taustar$, $\mustar$ and $\tnstar$ denote the sampled values. 
To generate $\taustar$, we draw $W_b \sim \chi^2_{k-1}$ and solve $\Q(\taus) = W_b$. Given $\taustar$, we obtain $\mustar$ as the solution of $p_\mathrm{E}(\mu \mid \taustar) = C(\mu \mid \taustar) = U_b$, where $U_b \sim \Uni(0,1)$. This exploits that the Edgington combined $p$-value function constitutes a confidence distribution function and is therefore uniformly distributed by the probability integral transform. All root-finding steps are implemented via Brent's algorithm using the \textsf{fntl} package \citep{Rfntl}. Finally, $\tnstar$ is drawn from $\Nor(\mustar, \taustar)$. Samples $\tau_{(1)}^{2*},\dots, \tau_{(B)}^{2*}$ are independent by independence of $W_1,\dots,W_B$, and samples $\mu_{(1)}^{*}, \dots, \mu_{(B)}^{*}$ are independent by independence of $U_1,\dots,U_B$ and of $W_b$ with $U_b$. Conditional on these, samples $\theta^{*(1)}_\text{new}, \dots, \theta^{*(B)}_\text{new}$ are independent, so that independence holds within each layer of the hierarchy. The Monte Carlo sampling procedures for the different predictive distributions are given in Algorithm~\ref{alg:mc_sampling}.
Incidentally, the samples $\mu_{(1)}^{*}, \dots, \mu_{(B)}^{*}$ obtained conditional on $\tau_{(1)}^{2*}, \dots, \tau_{(B)}^{2*}$ estimate a confidence distribution of $\mu$ with density
\[
c(\mu) = \int_{\mathbb{R}^+} c(\mu \mid \taus)\, c(\taus)\, \d\taus.
\]
We refer to this as the CD-Edgington estimator of $\mu$, which extends the classical Edgington estimator based on a fixed $\htaus$ \citep{heldpawelhofman2024}. Applying a change of variables to the generalized heterogeneity statistic $\Q(\taus)$ yields the analytic confidence density $c(\taus)$, which allows the estimator to be computed using adaptive quadrature integration. A discussion and comparison with the Monte Carlo estimator is provided in Web Appendix~B.

\begin{algorithm}[ht]
\renewcommand{\baselinestretch}{1.15}\selectfont
\caption{Monte Carlo sampling for predictive distributions in draw $b$.}
\label{alg:mc_sampling}
\begin{algorithmic}[1]
\Procedure{PCD-fixed}{$b$}
    \State Set $\taustar \gets \htaus$
    \State Draw $U_b \sim \Uni(0,1)$
    \State Set $\mustar \gets \text{root of } C(\mu \mid \htaus)=U_b$
    \State Draw $\tnstar \sim \Nor(\mustar,\htaus)$
\EndProcedure
\Procedure{PCD-simplified}{$b$}
    \State Draw $W_b \sim \chi^2_{k-1}$
    \State Set $\taustar \gets \text{root of } \Q(\taus)=W_b$
    \State Draw $U_b \sim \Uni(0,1)$
    \State Set $\mustar \gets \text{root of } C(\mu \mid \htaus)=U_b$
    \State Draw $\tnstar \sim \Nor(\mustar,\taustar)$
\EndProcedure
\Procedure{PCD-full}{$b$}
    \State Draw $W_b \sim \chi^2_{k-1}$
    \State Set $\taustar  \gets \text{root of } \Q(\taus)=W_b$
    \State Draw $U_b \sim \Uni(0,1)$
    \State Set $\mustar \gets \text{root of } C(\mu \mid \taustar)=U_b$
    \State Draw $\tnstar \sim \Nor(\mustar,\taustar)$
\EndProcedure
\end{algorithmic}
\end{algorithm}

We recommend generating at least $B = 100{,}000$ draws to reduce Monte Carlo noise. For three to 50 studies, this requires approximately four to eight seconds for the PCD-full distribution and yields relatively stable quantiles. The empirical distribution of $\tnstarB$ estimates the predictive distribution and enables computation of prediction intervals, including equi-tailed and highest confidence density (HCD) intervals. The HCD interval is defined analogously to a highest posterior density interval in Bayesian statistics, minimizing interval width for a given coverage:
\begin{definition}{Highest confidence density prediction interval:}
Let $\gamma \in (0,1)$ be a fixed prediction level. A $\gamma \times 100\%$ prediction interval $I = [c_l, c_u]$ is called a \emph{highest confidence density prediction interval} (HCDP interval) if $\forall \, \tn \in I \ \text{and} \ \widetilde{\theta}_{\scriptstyle{\text{new}}} \notin I$, $c(\tn) \ge c(\widetilde{\theta}_{\scriptstyle{\text{new}}})$.
\end{definition}
Equi-tailed and HCDP intervals may differ as the predictive distributions are not constrained to be symmetric, but inherit skewness from Edgington's confidence distribution. In intervention-related meta-analyses, it may be valuable to quantify the likelihood that a future study effect is beneficial, neutral, or harmful \citep{Riley2011}. We follow  \citet{Marschner2024} and refer to this as the \emph{confidence} or \emph{confidence probability} of the event to emphasize its epistemic interpretation. For example, the confidence that $\tn$ is equal to or larger than a margin $\delta$ is $\text{Conf}(\tn \ge \delta) = \int_{\delta}^{\infty} c(\tn) \d\tn \approx \tfrac{1}{B} \sum_{b = 1}^B \mathbf{1}\left\{\tnstar \ge \delta\right\}$, where $\mathbf{1}\{\cdot\}$ is the indicator function. The proposed methods are available in \textsf{R} through the \textsf{edgemeta} package (\url{https://github.com/davidkronthaler-dk/edgemeta}).

\subsection{Example: Corticosteroids and Mortality in Hospitalized COVID-19 Patients}
\label{sec:COVID19}

\citet{heldpawelhofman2024} considered a meta-analysis of seven randomized controlled trials reporting log odds ratios quantifying the association between corticosteroids, compared to placebo or standard care, and mortality in hospitalized COVID-19 patients \citep{who2020corticosteroids}. We revisit this example to illustrate the proposed predictive distributions and the CD-Edgington estimator. The study-level data are summarized in Web Table~\ref{tab:coviddata}.

\begin{figure}
\centering
\begin{knitrout}
\definecolor{shadecolor}{rgb}{0.969, 0.969, 0.969}\color{fgcolor}

{\centering \includegraphics[width=1\linewidth]{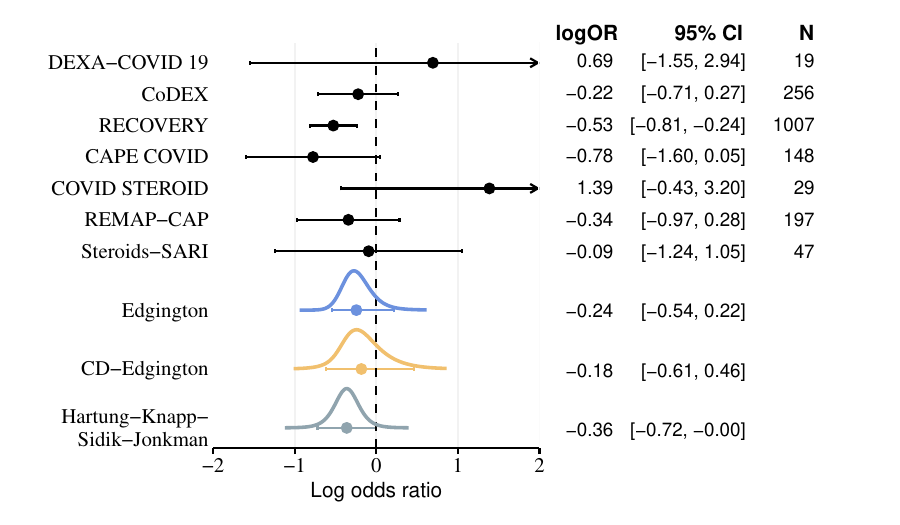} 

}

\end{knitrout}
\caption{Forest plot displaying results of a random-effects meta-analysis of seven randomized controlled trials reporting log odds ratios (logOR) quantifying the association between corticosteroids and mortality in hospitalized COVID-19 patients \citep{who2020corticosteroids}. Confidence densities of the average effect from the Edgington, CD-Edgington and Hartung--Knapp--Sidik--Jonkman method are displayed on top of 95\% confidence intervals (CI).}
\label{fig:forest}
\end{figure}

Figure~\ref{fig:forest} presents the results of a random-effects meta-analysis. Between-study heterogeneity is estimated using the Paule--Mandel estimator \citep{paule1982consensus} to match the approach of the original analysis in  \citet{who2020corticosteroids}, yielding $\htaus$ = 0.03 (95\% confidence interval by \citealp{Jackson2013} from 0.00 to 2.13; Higgins' $I^2$ = 14.01\%). Using the HKSJ method, the estimated average effect is -0.36 (95\% confidence interval from -0.72 to -0.004, p = 0.048, interval skewness of 0.00), while the CD-Edgington estimator yields -0.18 (95\% confidence interval from -0.61 to 0.46, p = 0.39, interval skewness of 0.20), both indicating a beneficial effect of corticosteroids on average. Interval skewness is computed as $\beta = (u + \ell - 2m)/(u - \ell)$ \citep{Groeneveld1984}, where $\ell$ and $u$ denote the lower and upper bounds of the interval, and $m$ the center, defined as the estimated average effect for confidence intervals and the median of the predictive distribution for prediction intervals. Note that the CD-Edgington 95\% confidence interval is considerably wider than that obtained using Edgington's method without accounting for heterogeneity uncertainty, which ranges from -0.54 to 0.22.

Figure~\ref{fig:forest} also displays the confidence densities of the average effect obtained from the HKSJ, Edgington and CD-Edgington approach. Unlike the symmetric HKSJ confidence distribution, the CD-Edgington distribution has a Fisher skewness coefficient (FiSC) of 0.90, reflecting the right-skewness of the study estimates. The CD-Edgington distribution estimates a confidence probability of 0.80 that the average effect is negative on the log odds scale (favorable to corticosteroids), to be compared with 0.98 under the HKSJ approach.

\begin{figure}
\centering
\begin{knitrout}
\definecolor{shadecolor}{rgb}{0.969, 0.969, 0.969}\color{fgcolor}

{\centering \includegraphics[width=\maxwidth]{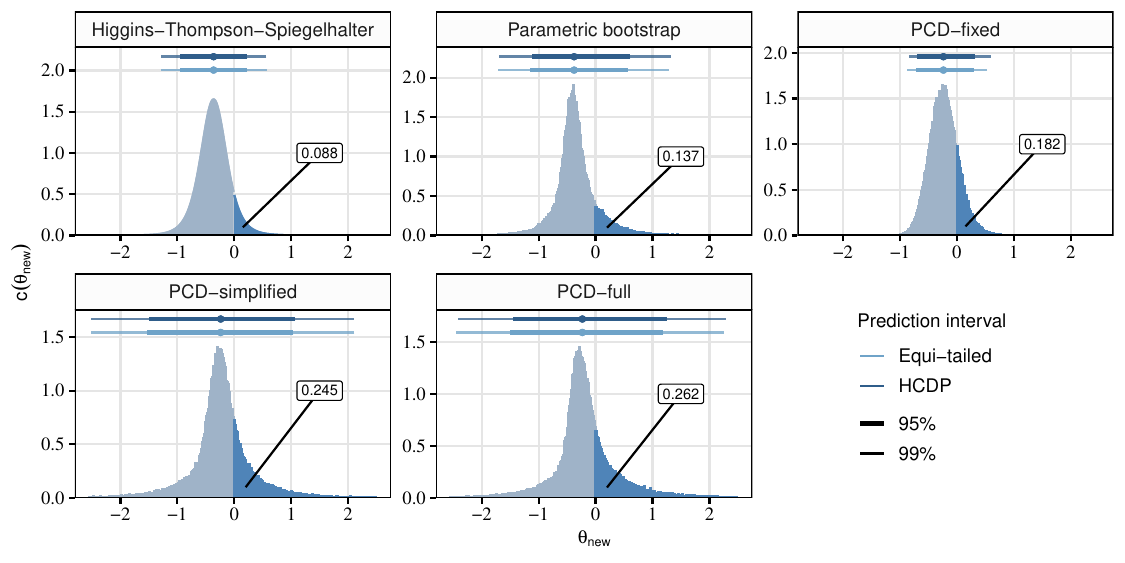} 

}

\end{knitrout}
\caption{Edgington's PCD and the predictive distributions of \citet{Higgins2008} and \citet{Nagashima2018}, estimated from seven randomized controlled trials reporting log odds ratios quantifying the association between corticosteroids and mortality in COVID-19 patients \citep{who2020corticosteroids}. Equi-tailed and highest confidence density prediction intervals (HCDP) at levels 95\% and 99\% are shown on top as telescope lines and confidence probabilities of future effects larger or equal to zero are indicated.}
\label{fig:sere_predictivedistributions}
\end{figure}

While Higgins' $I^2$ indicates that roughly 14\% of variability across study estimates is due to heterogeneity, it does not inform about absolute effect heterogeneity. In contrast, Figure~\ref{fig:sere_predictivedistributions}, which shows Edgington's PCD distributions alongside the predictive distributions by \citet{Higgins2008} and \citet{Nagashima2018}, provides direct estimates for the distribution of true effects. Displaying the entire predictive distribution, rather than only prediction intervals, allows for straightforward assessment of predictive information and uncertainty. Although the estimate of $\tau^2$ is close to zero, PCD-simplified and PCD-full remain relatively wide due to the substantial uncertainty in the heterogeneity estimate, which is reflected in the corresponding confidence distribution of $\tau^2$ (Web Figure~\ref{fig:covidconfdens}), over which these predictive distributions are obtained by marginalizing heterogeneity uncertainty. We observe that future effects are most often predicted to be beneficial concerning the treatment outcome, yet there is also a notable chance of effects in the opposite direction, even with large effect sizes for PCD distributions accounting for heterogeneity estimation uncertainty.

PCD-fixed is the narrowest, likely since it does not incorporate heterogeneity estimation uncertainty, and has FiSC = 0.42 with an estimated confidence probability that a future effect is zero or larger, $\widehat{\text{Conf}}(\tn \ge 0)$, of $0.182$, comparably larger than for the similarly dispersed but more left-shifted and symmetric HTS distribution ($0.088$). PCD-simplified and PCD-full are substantially wider, exhibiting $\widehat{\text{Conf}}(\tn \ge 0)$ of 0.24 and 0.26, respectively. Their corresponding FiSC's are 0.03 and 0.50, suggesting that while they inherit skewness from Edgington's confidence distribution, which has FiSC of 0.90, marginalization over the confidence density of $\taus$ reduces skewness, particularly for PCD-simplified in this example. The parametric bootstrap distribution is slightly narrower than PCD-simplifed and PCD-full, with FiSC of 0.37 and $\widehat{\text{Conf}}(\tn \ge 0) = 0.14$. Further, equi-tailed and HCDP intervals are similar, with differences mainly present for the PCD-fixed distribution. For simplicity we therefore use equi-tailed intervals throughout. In Web Tables~\ref{tab:covidpointestimates} and~\ref{tab:serepi}, we provide additional comparisons of 95\% prediction intervals and estimators of the average effect.

\section{Simulation Study}

\subsection{Design}

We present our phase~I/II simulation study \citep{heinze2024phases} according to the ADEMP framework \citep{morris2019using}.

\subsubsection{Aims}
Assess calibration and sharpness of the proposed predictive distributions and estimation properties of the proposed CD-Edgington estimator.

\subsubsection{Data-Generating Mechanism} 
The data-generating mechanism is adopted from \citet{heldpawelhofman2024} and described in detail in Web Appendix~D. We vary the number of studies $k \in$ \{3, 5, 10, 20, 50\}, the between-study heterogeneity determined by $\iota^2$ $\in$ \{0\%, 30\%, 60\%, 90\%\}, the number of large studies $k_{\text{large}} \in$ \{0, 1, 2\} and whether the true effects $\ti$ and future effects $\tn$ are generated from a normal distribution or from a left-skewed skew-normal distribution \citep{Azzalini2013}. Here, $\iota^2$ is the parametric counterpart to Higgins' $I^2$ \citep{higgins2025reflections}, defined via the variance decomposition $\tau^2 / (\tau^2 + \overline{\sigma}^2)$, where $\overline{\sigma}^2$ denotes the average within-study variance induced by the study sizes ($n_i = 50$ for normal and $n_i = 500$ for large studies). In each iteration, we generate $k$ true effects with effect estimates  and standard errors on the standardized mean difference scale, and 10{,}000 future effects $\tnj$, $j \in \{1,\dots,10{,}000\}$, to evaluate predictive performance. We perform the simulation study in a full-factorial manner.

\subsubsection{Estimands and Other Targets}
Predictive distributions estimate the joint effect distribution, corresponding to a $\Nor(\mu,\taus)$ with $\mu = -0.3$ and $\taus$ determined by $\iota^2$, or a skew-normal $\SN(\xi, \omega, \alpha)$ with skewness $\alpha = -4$ and location $\xi$ and scale $\omega$ obtained by moment-matching such that the mean is $-0.3$ and the variance is $\taus$. 

\subsubsection{Methods}
We compare equi-tailed 95\% prediction intervals from Edgington's PCD distributions with those from HTS \citep{Higgins2008} and the parametric bootstrap \citep{Nagashima2018}, the latter two accessed via the \textsf{meta} package \citep{Rmeta}. Performance measures based on the full predictive distribution are evaluated for the proposed and HTS methods; the full predictive distribution of \citet{Nagashima2018} is not available in implementations known to us. Evaluated estimators include Monte Carlo CD-Edgington, HKSJ, classical random-effects estimator, and Edgington's method without heterogeneity adjustment. Heterogeneity is estimated using REML for all methods, and stochastic methods (PCD, bootstrap, CD-Edgington) use $B = 100{,}000$ samples.

\subsubsection{Performance Measures}

The primary outcome is the coverage of 95\% prediction intervals. While nominal coverage is generally targeted, under no heterogeneity ($\iota^2 = 0\%$), 100\% coverage indicates that the interval covers the common effect $\mu$ which coincides with future effects $\tnj$. In each iteration, the estimated coverage is the proportion of future effects falling within the prediction interval \([I_\ell, I_u]\), i.e., $\frac{1}{10{,}000} \sum_{j = 1}^{10{,}000} \mathbf{1}\{ I_\ell \le \tnj \le I_u \}$. We run 4000 iterations per scenario, based on pilot studies to ensure a Monte Carlo error of the mean coverage of at most $0.005$ (Web Appendix~D).

Secondary outcomes are the width of 95\% prediction intervals, the Pearson correlation between the 95\% prediction interval skewness $\beta$ and the skewness of effect estimates, and the sign agreement quantified by Cohen's kappa. Effect estimate skewness is measured using Fisher's weighted skewness coefficient as in \citet{heldpawelhofman2024}. The HTS prediction interval, due to its additive construction always being symmetric, has always zero skewness.
Further, we compute continuous ranked probability scores (CRPS; \citealp{matheson1976scoring, gneiting2007strictly}), a strictly proper scoring rule that captures both calibration and sharpness. Lower values indicate better performance. The CRPS for a single observed future effect $\tnj$ is $\E\left[|\tnstarone - \tnj|\right] - \tfrac{1}{2}\E\left[|\tnstarone - \tnstartwo|\right]$, where $\tnstarone, \tnstartwo$ are two independent draws from the predictive distribution. Evaluating each predictive distribution against 10{,}000 future effects is computationally demanding, so we use a Monte Carlo approximation, where for a single future effect $\tnj$, the CRPS is approximated by $\frac{1}{B} \sum_{b = 1}^B | \tnstar - \tnj | - \frac{1}{2 (B - 1)} \sum_{b = 1}^{B-1} | \tnstar - \tnstarplus |$ \citep{gneiting2008assessing}. We report the mean CRPS across all future effects $\tnj$. Performance measures for estimation methods include 95\% confidence interval coverage and width, bias, root mean squared error, and skewness measures for confidence intervals analogous to those for prediction intervals.

\subsection{Results}

In 37 iterations ($0.008$\%) the bootstrap prediction interval did not converge. Non-convergences occurred only in scenarios with three studies and more frequently under large heterogeneity (Web Table~\ref{tab:ncnnf}). All other approaches always produced valid results. We applied listwise deletion and evaluated performance only based on convergent repetitions.

\subsubsection{Coverage of 95\% Prediction Intervals}

Figure~\ref{fig:picovernor} shows the mean coverage of 95\% prediction intervals for normally distributed effects. Without heterogeneity, all methods typically achieve mean coverage above the nominal level, as expected: under no heterogeneity, future effects coincide with the common effect, and prediction intervals that cover it achieve 100\% coverage. PCD-full and PCD-simplified intervals are closest to 100\% coverage, suggesting that the confidence distribution based on the generalized heterogeneity statistic accounts for heterogeneity even when point estimates are near zero.


Under heterogeneity, all methods approach nominal coverage as the number of studies increases, with generally better performance for larger heterogeneity. However, HTS and PCD-fixed intervals perform relatively poorly. For $\iota^2$ of 30\% and 60\%, the HTS interval shows a U-shaped pattern, over-covering with three studies and under-covering for five to 50 studies, though it performs well under large heterogeneity with at least five studies. The PCD-fixed interval consistently undercovers, often below 80\% with three studies, and approaches nominal coverage only with 50 studies and large heterogeneity.

Prediction intervals from PCD-full, PCD-simplified, and the parametric bootstrap generally achieve coverage close to nominal. Their performance is similar overall, with minor differences: the bootstrap is closest to nominal for $\iota^2$ of 30\% and 90\% with three studies, while PCD-full shows slightly better coverage in several scenarios with $\iota^2$ of 30\% and five to ten studies, and $\iota^2$ of 60\% and three studies. The PCD-simplified interval tends to deviate most from nominal coverage in scenarios with few studies among these three methods. PCD-full typically achieves higher coverage than PCD-simplified, reflecting the additional uncertainty incorporated.

Results under skew-normal effects are highly comparable, with only minor declines in coverage for these three well-performing methods (Web Figure~\ref{fig:picoversn}). Coverage distributions (Web Figures~\ref{fig:coverdist_nor0}-- \ref{fig:coverdist_sn90}) also align closely across methods and effect distributions for these approaches. In contrast, HTS often exhibits bimodal coverage distributions under $\iota^2$ of 30\% and 60\%, with modes near the nominal level and around 0.5. Similar but more dispersed patterns are observed for the PCD-fixed interval.

\afterpage{
\clearpage
\begin{landscape}
\begin{figure}
\centering
\begin{knitrout}
\definecolor{shadecolor}{rgb}{0.969, 0.969, 0.969}\color{fgcolor}

{\centering \includegraphics[width=1\linewidth]{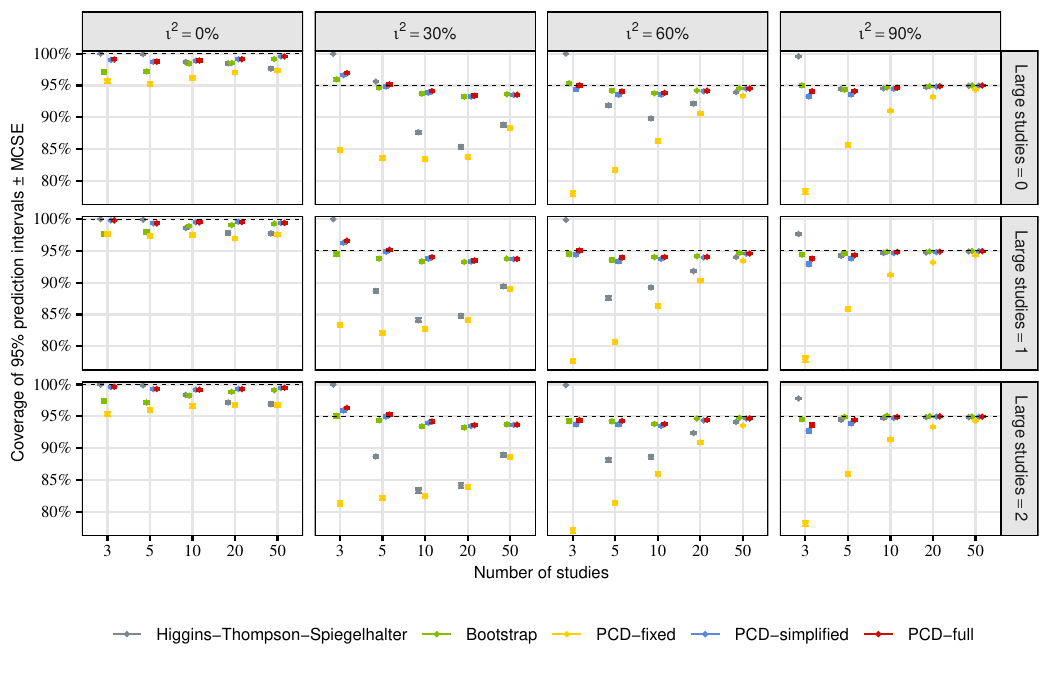} 

}

\end{knitrout}
\caption{Coverage of 95\% prediction intervals for true effects following a normal distribution. Error bars represent Monte Carlo standard errors (MCSE). For no heterogeneity ($\iota^2 = 0\%$), future effects coincide with the common effect, with 100\% coverage reflecting that prediction intervals cover the common effect.}
\label{fig:picovernor}
\end{figure}
\end{landscape}
\clearpage
}

\subsubsection{Prediction Interval Characteristics and CRPS}

Figure~\ref{fig:piskewhesnor} shows the Pearson correlation between the skewness of prediction intervals and effect estimates under normally distributed effects. Correlations are typically moderate to strong for PCD-full with fewer than 50 studies, and often for PCD-fixed with fewer than 20 studies, while PCD-simplified generally shows weaker correlations. As the number of studies increases, correlations for all methods approach zero, potentially reflecting increasing symmetry as estimates converge to the underlying normal distribution. The bootstrap interval is strongly influenced by the number of large studies: correlations remain near zero when none are present, but increase with more large studies, reaching levels comparable to the most skew-sensitive methods in settings with few studies and low to moderate heterogeneity. Similar patterns are observed under skew-normal effects (Web Figure~\ref{fig:piskewhessn}) and for Cohen's kappa (Web Figures~\ref{fig:pikappanor}, \ref{fig:pikappasn}).

\afterpage{
\clearpage
\begin{landscape}
\begin{figure}
\centering
\begin{knitrout}
\definecolor{shadecolor}{rgb}{0.969, 0.969, 0.969}\color{fgcolor}

{\centering \includegraphics[width=1\linewidth]{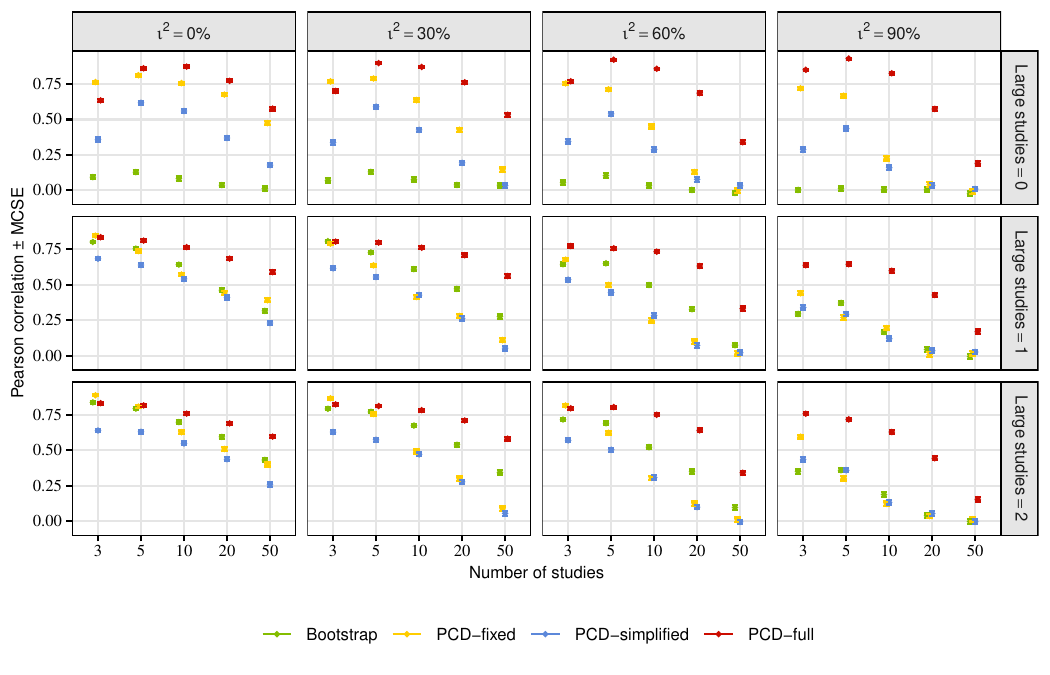} 

}

\end{knitrout}
\caption{Pearson correlation between the skewness of 95\% prediction intervals and the skewness of effect estimates for true effects following a normal distribution. Error bars represent Monte Carlo standard errors (MCSE).}
\label{fig:piskewhesnor}
\end{figure}
\end{landscape}
\clearpage
}

Web Figures~\ref{fig:piwnor} and~\ref{fig:piwsn} show the average width of 95\% prediction intervals. Trends are highly comparable across effect distributions. With three studies, the HTS interval is substantially wider than all others, especially under large heterogeneity, likely due to the heavy tails of the $t$-distribution at low degrees of freedom. The PCD-fixed interval is consistently the narrowest. PCD-full, PCD-simplified, and bootstrap intervals have similar widths across scenarios, although the bootstrap tends to be slightly wider with few studies. As the number of studies increases to 50, widths converge across all methods. The presence of large studies substantially narrows intervals for few studies.

Web Figures~\ref{fig:crpsnor} and~\ref{fig:crpssn} show the average CRPS, with highly comparable results across effect distributions. CRPS increases with heterogeneity, reflecting reduced sharpness. Predictive distributions based on Edgington's method outperform HTS when the number of studies is small and/or heterogeneity is low. For ten or more studies, CRPS is nearly identical across methods. With two large studies and $\iota^2$ of 60\% and 90\%, CRPS for Edgington-based methods increases with the number of studies, though the effect is modest and stabilizes from ten studies onward.

\subsection{Summary of Simulation Results for Prediction}

Simulation results indicate that PCD-full and PCD-simplified, which account for heterogeneity estimation uncertainty, are typically well calibrated for more than three studies, with 95\% prediction intervals often approaching nominal coverage. They often outperform HTS in coverage and perform comparably to the bootstrap interval, while sometimes yielding narrower intervals for few studies. The PCD-full interval captures skewness effectively, showing substantial correlations between interval and study estimate skewness with fewer than 50 studies. In contrast, the simpler PCD-fixed distribution, which ignores heterogeneity estimation uncertainty, approaches nominal coverage only for very large numbers of studies and is unlikely to be useful in practical applications.

\subsection{Summary of Simulation Results for Estimation}

The CD-Edgington estimator remains unbiased under normal effects (Web Figure~\ref{fig:biasnor}), with its 95\% confidence interval approaching nominal coverage in most scenarios with more than three studies and heterogeneity (Web Figure~\ref{fig:cicovernor}). Under no heterogeneity or with only three studies, intervals tend to overcover, though they are often narrower than HKSJ intervals (Web Figures~\ref{fig:ciwnor},~\ref{fig:ciwsn}). This overcoverage is expected, since CD-Edgington always propagates uncertainty in the heterogeneity parameter, whereas HKSJ can even lead to narrower confidence intervals than the random effects method (which conditions on the heterogeneity parameter) in the case of homogeneity \citep{Wiksten2016}. Under skew-normal effects, coverage decreases slightly and small bias appears under large heterogeneity (Web Figures~\ref{fig:biassn},~\ref{fig:cicoversn}). The CD-Edgington interval is generally wider than the version that ignores heterogeneity estimation uncertainty.

The HKSJ confidence interval typically achieves nominal coverage when study sample sizes are equal, though this is rarely the case in practice. Already noted by \citet{IntHout2014}, with unequal study sizes, HKSJ may undercover for $\iota^2 = 30$\% with $\leq 20$ studies, $\iota^2 = 60$\% with $\leq 10$ studies, and $\iota^2 = 90$\% with three studies. Despite this, HKSJ is recommended in the Cochrane Handbook when heterogeneity is estimated greater zero and more than two studies are available \citep{deeks2024chapter10}, which may require clarification regarding the influence of study sizes. Additionally, with three studies, HKSJ intervals are typically very wide due to the heavy-tailed $t_1$-distribution. In contrast, CD-Edgington intervals are less sensitive to study sample sizes, with coverage typically closer to nominal under heterogeneity and one or two large studies. 

\section{Discussion}

We proposed a novel approach for constructing predictive distributions in random-effects meta-analysis that accounts for parameter uncertainty through Edgington's confidence distribution and through a confidence distribution of the heterogeneity parameter. In practice, predictive distributions are commonly summarized by prediction intervals, partly because many widely used software implementations provide only such summaries. However, we consider this reduction unnecessary for many predictive tasks, as it discards distributional information. By enabling direct access to samples from the predictive distribution, our method facilitates broader inferences beyond interval summaries.

We applied the confidence distribution of the generalized heterogeneity statistic, which varies in $\taus$, to incorporate uncertainty with respect to the heterogeneity parameter, while \citet{Nagashima2018} employed the confidence distribution of the standard $\Q$ statistic, whose exact distribution has parameters dependent on $\taus$. The method of \citet{Nagashima2018} has been evaluated across several independent simulation studies (including ours, \citealp{Nagashima2018}, and \citealp{matrai_archive}) and has demonstrated good performance, consistently outperforming the HTS method in terms of coverage in many settings.

Results from our simulation study suggest that the proposed methods perform similarly to the bootstrap interval in terms of calibration and sharpness, and likewise often show improved performance compared to the HTS approach in the investigated settings. While the bootstrap method currently has stronger empirical support, our method would benefit from further independent evaluation, for example across alternative effect scales beyond standardized mean differences or under alternative effect distributions, as in \citet{matrai_archive}. Further, following \citet{held2024assessment}, Edgington's predictive distributions can be extended by incorporating study-specific weights to downweight studies at risk of bias. The combined $p$-value function \eqref{eq:pEdgington} has to be extended to include weights, as described in Web Appendix~F. Applying inverse-standard-error and inverse-variance weights to the example presented in Section \ref{sec:COVID19} yields point estimates and predictions that are now closer to the results from the largest study (RECOVERY), but still have considerable uncertainty. More research is needed to investigate operating characteristics of Edgington predictions intervals with different weights in a simulation study.

Since the proposed predictive distributions are restricted to be unimodal and are based on a normal random-effects model, future work could incorporate more flexible effect distributions. For instance, skew-normal models have previously been applied in Bayesian meta-analysis \citep{Lee2007} and in diagnostic test accuracy meta-analysis under a frequentist framework \citep{Negeri2020}, with both studies reporting improved precision when effects are truly skewed. Our methodology could be extended by using standardized skew-normal pivots to construct study specific $p$-value functions, conditional on moment-estimated shape and scale parameters \citep{Thiuthad2019} under a univariate formulation of the normal-skew-normal model by \citet{Negeri2020}. A confidence distribution of the location parameter could then be constructed by combining $p$-values using Edgington's method or another $p$-value combination technique. However, it remains unclear how to appropriately incorporate uncertainty about the scale and shape parameter estimates, which we strongly recommend, as we here have demonstrated the value of accounting for estimation uncertainty in meta-analytic models.

In summary, we have discussed Edgington's approach for combining $p$-value functions, or equivalently, confidence distributions, for prediction and estimation in random-effects meta-analysis. Although further investigation is needed, results indicate that Edgington's method has potential as a viable alternative or complement to classical random-effects meta-analysis. 

\newpage
\bibliographystyle{apalike}
\bibliography{../paper-ref}


\renewcommand{\thetable}{\arabic{table}}
\renewcommand{\thefigure}{\arabic{figure}}

\newcommand{\hl}{\textcolor{red}}

\makeatletter
\renewcommand{\fnum@table}{Web Table \thetable}
\renewcommand{\fnum@figure}{Web Figure \thefigure}
\makeatother

\clearpage
\setcounter{page}{1}
\renewcommand{\thepage}{S\arabic{page}}

\begin{titlepage}
\centering
\vspace*{2cm}
{\LARGE\bfseries Supplementary Materials\par}
\vspace{1.5em}
{\large for\par}
\vspace{1.5em}
{\LARGE\bfseries Prediction intervals for random-effects meta-analysis based on confidence distributions and Edgington's method\par}
\vfill
\end{titlepage}
\clearpage

\setcounter{table}{0}
\setcounter{figure}{0}

\section*{Web Appendix A: \textit{P}-Value Functions and Confidence Distributions: Example and Applications}

\begin{figure}[h!]
\centering
\begin{knitrout}
\definecolor{shadecolor}{rgb}{0.969, 0.969, 0.969}\color{fgcolor}

{\centering \includegraphics[width=1\linewidth]{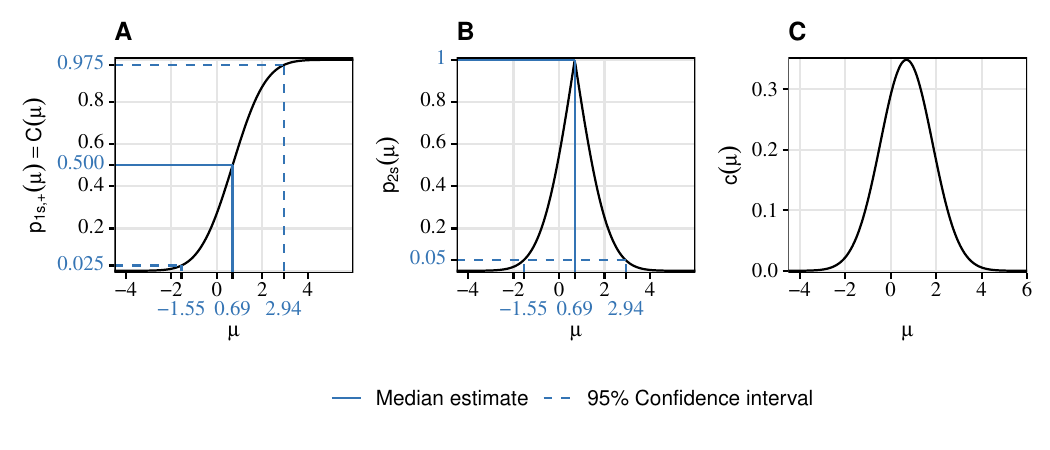} 

}

\end{knitrout}
\caption{Results from the DEXA-COVID-19 trial (NCT04325061), see Web Table~\ref{tab:coviddata}. Wald test for $\mu$ with $\hmu = -0.69$ and $\se(\hmu) = 1.15$: (A) One-sided $p$-value function for the alternative "greater", corresponding to the confidence distribution function; (B) two-sided $p$-value function; (C) confidence density.}
\label{fig:pfunexample}
\end{figure}

Let 
\[
Z(\mu) = \frac{\hmu - \mu}{\se(\hmu)}
\]
denote the standard normal pivot of the Wald test for a parameter $\mu$ with estimate $\hmu$. The corresponding one-sided (1s) and two-sided (2s) $p$-value functions are given by
\begin{align*}
p_{\text{1s},+}(\mu) &= 1 - \Phi\left\{Z(\mu)\right\} && \text{for the alternative "greater"}, \\
p_{\text{1s},-}(\mu) &= \Phi\left\{Z(\mu)\right\} && \text{for the alternative "less"}, \\
p_{\text{2s}}(\mu) &= 2 \min\left\{p_{\text{1s},+}(\mu),\; p_{\text{1s},-}(\mu)\right\},
\end{align*}
where $\Phi$ denotes the standard normal cumulative distribution function (CDF; \citealp{Fraser2019}). In the meta-analysis case study presented in Section~2.6 of the main manuscript, the included DEXA-COVID-19 trial (NCT04325061) reports an estimated log odds ratio quantifying the association between corticosteroids and mortality in hospitalized COVID-19 patients of $\hmu = 0.69$ with standard error $\se(\hmu) = 1.15$. Web Figures~\ref{fig:pfunexample}A and~\ref{fig:pfunexample}B display the one-sided $p$-value function for the ``greater'' alternative and the two-sided $p$-value function for the Wald test, respectively.

Point and interval estimates can be directly obtained from the $p$-value functions. The median estimator for $\mu$ satisfies
\[
\hmu_\mathrm{Med} = p_{\text{1s},+}^{-1}(0.5) = p_{\text{1s},-}^{-1}(0.5),
\]
yielding $\hmu_\mathrm{Med} = 0.69$ in the DEXA-COVID-19 trial. Similarly, the median estimate is given by the value that maximizes the two-sided $p$-value function. A two-sided $(1-\alpha) \times 100\%$ confidence interval for $\mu$, obtained from the one-sided $p$-value functions, ranges from
\[
p_{\text{1s},+}^{-1}\!\left(\tfrac{\alpha}{2}\right) 
= p_{\text{1s},-}^{-1}\!\left(1 - \tfrac{\alpha}{2}\right)
\quad \text{to} \quad
p_{\text{1s},+}^{-1}\!\left(1 - \tfrac{\alpha}{2}\right) 
= p_{\text{1s},-}^{-1}\!\left(\tfrac{\alpha}{2}\right).
\]
For example, the two-sided 95\% confidence interval ($\alpha = 0.05$) for the DEXA-COVID-19 trial ranges from $-1.55$ to $2.94$. Equivalently, the same interval can be obtained from the two-sided $p$-value function by identifying the set of parameter values for which $p_{\text{2s}}(\mu) \geq \alpha = 0.05$, i.e., by intersecting the two-sided $p$-value function with a horizontal line at $\alpha = 0.05$.

While these quantities are straightforward to obtain in the single-study setting, where they coincide with the usual Wald estimator and confidence interval, the principle extends to meta-analysis through the use of combined $p$-value functions, as investigated by \citet{heldpawelhofman2024}. In particular, point and interval estimates can be obtained from the combined $p$-value function in an analogous manner.

Further, the value of the one-sided $p$-value function at any point in the parameter space corresponds to the $p$-value for testing the corresponding one-sided null hypothesis. Specifically,
\begin{align*}
p_{\text{1s},+}(\mu_0) 
&\text{ is the $p$-value for testing } H_0:\mu \le \mu_0 \text{ vs. } H_1:\mu > \mu_0 \ \text{and} \\[4pt]
p_{\text{1s},-}(\mu_0) 
&\text{ is the $p$-value for testing } H_0:\mu \ge \mu_0 \text{ vs. } H_1:\mu < \mu_0.
\end{align*}
Similarly, $p_{\text{2s}}(\mu_0)$ is the $p$-value for testing $H_0:\mu = \mu_0$ against the two-sided alternative $H_1:\mu \neq \mu_0$. Hence, the $p$-value function provides a complete mapping from the parameter space to the corresponding $p$-values of hypothesis tests, thereby enabling simultaneous inference for all possible null hypotheses.

Moreover, the one-sided $p$-value function for the ``greater'' alternative typically coincides with a confidence distribution function of the parameter, with exceptions arising, for example, in strictly conservative one-sided tests (see Section~2.3.2 in \citealp{Xie2013}). Consequently, Web Figure~\ref{fig:pfunexample}A can also be interpreted as the confidence distribution function of $\mu$. Its derivative with respect to $\mu$, the confidence density, is shown in Web Figure~\ref{fig:pfunexample}C.

The modern interpretation of a confidence distribution is that it provides a distributional summary of uncertainty about the parameter, rather than representing an intrinsic probability distribution of the parameter itself \citep{Xie2013}. Intuitively, it can be viewed as the collection of all one-sided (left-sided) confidence intervals. In particular, the value of the confidence distribution function at $\mu$, denoted by $C(\mu)$, equals the largest confidence level $1-\alpha$ for which $\mu$ is contained in the corresponding left-sided confidence interval, that is,
\[
C(\mu) = \sup \left\{ 1 - \alpha : \mu \in (-\infty,\, u_{1-\alpha}] \right\},
\]
where $(-\infty,\, u_{1-\alpha}]$ denotes the left-sided $(1-\alpha)\times 100\%$ confidence interval. For more detailed discussions of $p$-value functions and confidence distributions, we refer the interested reader to \citet{Xie2013, Schweder2016, infanger2019p, Fraser2019, Marschner2024}.

\clearpage
\section*{Web Appendix B: The CD-Edgington estimator}

\subsection*{B.1 Monte Carlo Construction of the CD-Edgington Estimator}

We employ a Monte Carlo sampling algorithm to generate draws $(\taustar, \mustar, \tnstar)$, $b = 1, \dots, B$, in order to construct the proposed predictive distributions (Algorithm~\ref{alg:mc_sampling} in the main text). For the PCD-full procedure, each draw $b$ is obtained as follows: first, a value $\taustar$ is sampled from the confidence distribution of $\taus$, implied by the generalized heterogeneity statistic \citep{Viechtbauer2007}. Conditional on this draw, a value $\mustar$ is generated from Edgington's confidence distribution, using that, by the probability integral transform, it follows a standard uniform distribution.

The resulting samples $\mustar$, obtained conditionally on $\taustar$, provide an empirical estimator of a confidence distribution of $\mu$ of the form
\begin{equation}\label{eq:cdedge}
c(\mu) = \int_{\mathbb{R}^+} c(\mu \mid \taus)\, c(\taus)\, d\taus.
\end{equation}
Equi-tailed confidence intervals are obtained from the corresponding empirical quantiles of the Monte Carlo samples, while point estimates are obtained as the empirical means. This extends the estimator of \citet{heldpawelhofman2024}, which is based on $c(\mu \mid \taus)$ with a plug-in estimate $\htaus$ for $\taus$. In contrast, our approach incorporates uncertainty in the heterogeneity parameter $\taus$ by integrating over its confidence distribution. We refer to this as the CD-Edgington estimator.

Importantly, the CD-Edgington estimator assigns non-zero mass to values $\taus > 0$ through the confidence distribution of $\taus$, even when point estimates may suggest negligible heterogeneity. Consequently, it is inherently a random-effects approach and is not applicable in a fixed-effect framework.

\subsection*{B.2 Global Adaptive Quadrature for the CD-Edgington Estimator}

While this estimator arises naturally from the Monte Carlo sampling, the integral can alternatively be evaluated using deterministic numerical integration. In particular, we employ a global adaptive quadrature (GAQ) algorithm \citep{piessens2012quadpack, Rfntl}. The GAQ approach is based on a change of variables argument applied to the generalized heterogeneity statistic $\Q(\taus)$ to obtain the confidence density of $\taus$. Since $\Q(\taus)$ is monotonically decreasing in $\taus$ and differentiable, and noting that $\taus = \Q^{-1}\{\Q(\taus)\}$, the confidence density of $\taus$ is given by
\[
c(\taus) = f_{\chi^2_{k-1}}\!\left\{\Q(\taus)\right\}
\left| \frac{\d \Q(\taus)}{\d \taus} \right|,
\]
where $f_{\chi^2_{k-1}}$ denotes the density of a $\chi_{k-1}^2$-distribution. Recall that the generalized heterogeneity statistic is
\[
\Q(\tau^2) = \sum_{i=1}^k w_i(\tau^2)
\big\{\widehat{\theta}_i - \widehat{\mu}_{\mathrm{IVW}}(\tau^2)\big\}^2,
\]
where
\[
w_i(\tau^2) = \big\{\tau^2 + \mathrm{se}(\widehat{\theta}_i)^2\big\}^{-1} \qquad \text{and} \qquad 
\widehat{\mu}_{\mathrm{IVW}}(\tau^2) = \frac{\sum_{i=1}^k w_i(\tau^2)\widehat{\theta}_i}{\sum_{i=1}^k w_i(\tau^2)}
\]
is the inverse-variance weights random-effects estimator as a function of $\tau^2$. Differentiating $\Q(\taus)$ using the product rule yields
\[
\frac{\d\Q(\taus)}{\d\taus}
= \sum_{i=1}^k \left[
\frac{\d \wit}{\d \taus} \big\{\hti - \hmuivw(\taus)\big\}^2
- 2\,\wit\big\{\hti - \hmuivw(\taus)\big\}\frac{\d \hmuivw(\taus)}{\d \taus}
\right],
\]
with
\[
\frac{\d \wit}{\d \taus} = -\frac{1}{(\taus + \hsi)^2}.
\]
The derivative of $\hmuivw(\taus)$ follows from the quotient rule:
\[
\frac{\d \hmuivw(\taus)}{\d \taus}
=
\frac{
\left\{\sum_{i=1}^k \wit\right\}
\left\{\sum_{i=1}^k \frac{\d \wit}{\d \taus}\,\hti\right\}
-
\left\{\sum_{i=1}^k \wit\,\hti\right\}
\left\{\sum_{i=1}^k \frac{\d \wit}{\d \taus}\right\}
}{
\left\{\sum_{i=1}^k \wit\right\}^2
}.
\]
Substituion yields
\[
\frac{\d \Q(\taus)}{\d\taus}
= \sum_{i=1}^k \left[
- \frac{\big\{\hti - \hmuivw(\taus)\big\}^2}{(\taus + \hsi)^2}
+ \frac{2}{\taus + \hsi}\big\{\hti - \hmuivw(\taus)\big\}
\frac{\d \hmuivw(\taus)}{\d\taus}
\right],
\]
which allows to construct the analytic confidence density $c(\taus)$. Using this analytic confidence density, the integral in~\eqref{eq:cdedge} is evaluated numerically via global GAQ. Equi-tailed confidence intervals are obtained by inverting the corresponding CDF, while point estimates are computed by approximating the expected value using a weighted sum over interval midpoints based on finite differences of the CDF.

\subsection*{B.3 Illustrative Examples of Heterogeneity Confidence Densities}

\begin{figure}
\centering
\begin{knitrout}
\definecolor{shadecolor}{rgb}{0.969, 0.969, 0.969}\color{fgcolor}

{\centering \includegraphics[width=0.9\linewidth]{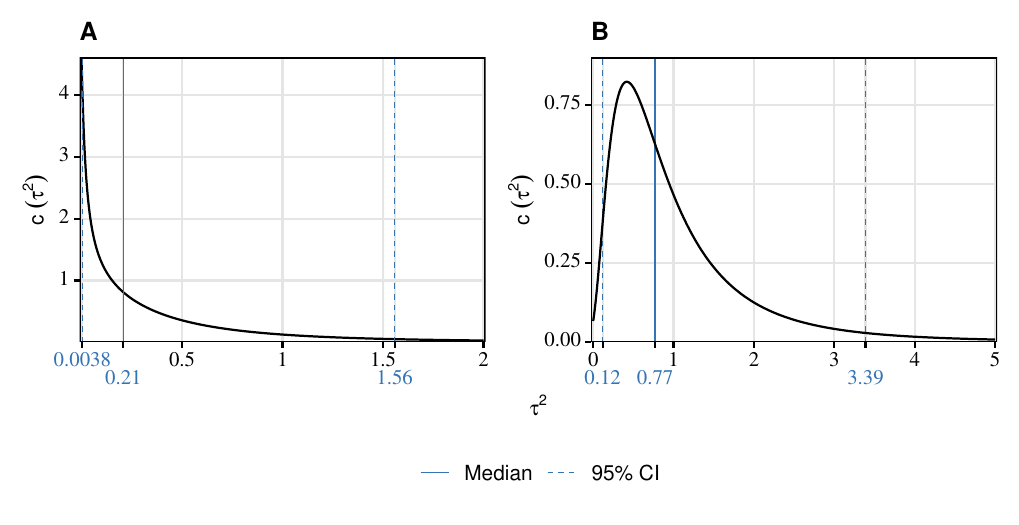} 

}

\end{knitrout}
\caption{Confidence densities of the between-study heterogeneity parameter $\taus$, based on the generalized heterogeneity statistic. Panel~A corresponds to seven reported log odds ratios quantifying the association between corticosteroids and mortality in hospitalized COVID-19 patients \citep{who2020corticosteroids}, while Panel~B corresponds to nine reported mean differences on \textit{Serenoa repens} treatment for urinary tract symptoms \citep{Franco2023}.}
\label{fig:cdtau2}
\end{figure}

Web Figure~\ref{fig:cdtau2} displays the confidence densities of the heterogeneity parameters for two meta-analyses. The first corresponds to the case study presented in the main manuscript, synthesizing seven reported log odds ratios quantifying the association between corticosteroids and mortality in hospitalized COVID-19 patients \citep{who2020corticosteroids}, with an estimated $\htaus = 0.03$ (95\% confidence interval by  \citealp{Jackson2013} from 0.00 to 2.13; Higgins' $I^2$ = 14.01\%), estimated using the Paule--Mandel approach \citep{paule1982consensus}. The second example is based on nine reported mean differences investigating the effect of \textit{Serenoa repens} treatment on lower urinary tract symptoms \citep{Franco2023}, yielding $\htaus$ of 0.68 (95\% confidence interval by  \citealp{Jackson2013} from 0.11 to 3.96; Higgins' $I^2$ = 67.36\%). 

In the first example, where heterogeneity is small, the confidence density is sharply peaked at zero. In contrast, the second example with substantial heterogeneity produces a broader distribution with its peak away from zero. The median estimates for $\taus$ obtained from the confidence distribution approach presented here are 0.21 (95\% confidence interval from 0.00 to 1.56) and 0.77 (95\% confidence interval from 0.12 to 3.39), respectively. While these summaries are provided for comparison with Paule--Mandel estimates, we emphasize that this approach is not intended to reduce the confidence distribution of $\taus$ to a scalar summary or interval, but rather to represent uncertainty through the full confidence distribution.

\subsection*{B.4 Comparison of Monte Carlo and GAQ Integration}

Web Tables~\ref{tab:MCvsGAQ} and~\ref{tab:MCvsGAQ_bc} display the results from a pilot simulation with 1000 iterations, varying the number of studies  $k \in \{3,5,10,20,50\}$ and between-study heterogeneity $\iota^2 \in \{0\%, 30\%, 60\%, 90\%\}$ for normally distributed true effects and no large studies according to the simulation design presented in the main manuscript, described in detail in Web Appendix~D below. 

Web Table~\ref{tab:MCvsGAQ} displays mean differences in point estimates and 95\% confidence interval limits between the two integration approaches. Web Table~\ref{tab:MCvsGAQ_bc} shows bias of point estimators and coverage of 95\% confidence intervals for both methods. We found that the GAQ approach tends to produce slightly too wide marginal distributions and confidence intervals in scenarios with three or five studies. For ten or more studies, differences in confidence interval limits could be largely attributed to Monte Carlo error. Point estimates from both approaches were nearly identical. Although GAQ offers greater computational efficiency and avoids Monte Carlo noise, the results suggest numerical instability under scenarios with few studies, which may impact performance. Hence, we recommend using the Monte Carlo algorithm for computing confidence intervals, particularly in meta-analyses with few studies.

\subsection*{B.5 Reconstruction of Combined \textit{P}-Value Functions}

Both approaches allow us to reconstruct the combined $p$-value function, e.g. to display in drappery plots \citep{rucker2021beyond}, since the upper (respectively, lower) tail mass of a confidence distribution at $\mu_0$ equals the $p$-value of the one-sided test for $H_0: \mu \le \mu_0$ against $H_1: \mu > \mu_0$ (respectively, $H_0: \mu \ge \mu_0$ against $H_1: \mu < \mu_0$; \citealp{Xie2013}). For the Monte Carlo algorithm, this can be achieved by considering the empirical CDF, whereas the GAQ approach requires additional integration steps.

\begin{table}[ht]
\centering
\caption{Mean differences (with Monte Carlo standard errors) in point estimates and 95\% confidence interval limits between Monte Carlo sampling and global adaptive quadrature integration for the CD-Edgington estimator.} 
\label{tab:MCvsGAQ}
\begin{tabular}{lrrrr}
   \midrule
Estimate (MC - GAQ) &  &  &  &  \\ 
  Studies ($k$) & $\iota^2$ = 0\% & 30\% & 60\% & 90\% \\ 
   \midrule
3 & -0.000 [0.000] & -0.000 [0.000] & -0.000 [0.000] & -0.000 [0.000] \\ 
  5 & 0.000 [0.000] & 0.000 [0.000] & -0.000 [0.000] & -0.000 [0.000] \\ 
  10 & -0.000 [0.000] & -0.000 [0.000] & 0.000 [0.000] & 0.000 [0.000] \\ 
  20 & -0.000 [0.000] & -0.000 [0.000] & 0.000 [0.000] & 0.000 [0.000] \\ 
  50 & 0.000 [0.000] & -0.000 [0.000] & 0.000 [0.000] & -0.000 [0.000] \\ 
   \midrule
Lower 95\% CI (MC - GAQ) &  &  &  &  \\ 
  Studies ($k$) & $\iota^2$ = 0\% & 30\% & 60\% & 90\% \\ 
  3 & 0.069 [0.003] & 0.056 [0.003] & 0.037 [0.003] & -0.088 [0.005] \\ 
  5 & 0.030 [0.001] & 0.029 [0.001] & 0.025 [0.001] & 0.010 [0.000] \\ 
  10 & 0.009 [0.000] & 0.009 [0.000] & 0.008 [0.000] & 0.004 [0.000] \\ 
  20 & 0.004 [0.000] & 0.004 [0.000] & 0.004 [0.000] & 0.002 [0.000] \\ 
  50 & 0.002 [0.000] & 0.002 [0.000] & 0.002 [0.000] & 0.002 [0.000] \\ 
   \midrule
Upper 95\% CI (MC - GAQ) &  &  &  &  \\ 
  Studies ($k$) & $\iota^2$ = 0\% & 30\% & 60\% & 90\% \\ 
  3 & -0.069 [0.003] & -0.056 [0.003] & -0.037 [0.003] & 0.088 [0.005] \\ 
  5 & -0.030 [0.001] & -0.029 [0.001] & -0.025 [0.001] & -0.010 [0.000] \\ 
  10 & -0.009 [0.000] & -0.009 [0.000] & -0.008 [0.000] & -0.004 [0.000] \\ 
  20 & -0.004 [0.000] & -0.004 [0.000] & -0.004 [0.000] & -0.002 [0.000] \\ 
  50 & -0.002 [0.000] & -0.002 [0.000] & -0.002 [0.000] & -0.002 [0.000] \\ 
    
\multicolumn{5}{l}{\footnotesize CI = confidence interval, GAQ = global adaptive quadrature, MC = Monte Carlo.} \\ 
 \bottomrule
\end{tabular}
\end{table}

\begin{table}[ht]
\centering
\caption{Bias of point estimators and coverage of 95\% confidence intervals (with Monte Carlo standard errors) for Monte Carlo sampling and global adaptive quadrature integration for the CD-Edgington estimator.} 
\label{tab:MCvsGAQ_bc}
\begin{tabular}{lrrrr}
   \midrule
MC: Bias &  &  &  &  \\ 
  Studies ($k$) & $\iota^2$ = 0\% & 30\% & 60\% & 90\% \\ 
   \midrule
3 & 0.004 [0.004] & -0.001 [0.004] & -0.001 [0.005] & -0.007 [0.007] \\ 
  5 & 0.000 [0.003] & 0.001 [0.003] & 0.001 [0.003] & 0.003 [0.005] \\ 
  10 & -0.002 [0.002] & -0.002 [0.002] & -0.002 [0.002] & -0.003 [0.003] \\ 
  20 & 0.002 [0.001] & -0.001 [0.001] & 0.001 [0.001] & 0.002 [0.002] \\ 
  50 & 0.001 [0.001] & -0.000 [0.001] & 0.000 [0.001] & -0.000 [0.001] \\ 
   \midrule
GAQ: Bias &  &  &  &  \\ 
  Studies ($k$) & $\iota^2$ = 0\% & 30\% & 60\% & 90\% \\ 
  3 & 0.004 [0.004] & -0.001 [0.004] & -0.000 [0.005] & -0.007 [0.007] \\ 
  5 & 0.000 [0.003] & 0.001 [0.003] & 0.001 [0.003] & 0.003 [0.005] \\ 
  10 & -0.001 [0.002] & -0.002 [0.002] & -0.002 [0.002] & -0.003 [0.003] \\ 
  20 & 0.002 [0.001] & -0.001 [0.001] & 0.001 [0.001] & 0.002 [0.002] \\ 
  50 & 0.001 [0.001] & -0.000 [0.001] & 0.000 [0.001] & -0.000 [0.001] \\ 
   \midrule
MC: 95\% CI coverage &  &  &  &  \\ 
  Studies ($k$) & $\iota^2$ = 0\% & 30\% & 60\% & 90\% \\ 
  3 & 0.980 [0.004] & 0.981 [0.004] & 0.971 [0.005] & 0.960 [0.006] \\ 
  5 & 0.973 [0.005] & 0.965 [0.006] & 0.959 [0.006] & 0.940 [0.008] \\ 
  10 & 0.963 [0.006] & 0.962 [0.006] & 0.957 [0.006] & 0.960 [0.006] \\ 
  20 & 0.966 [0.006] & 0.961 [0.006] & 0.952 [0.007] & 0.951 [0.007] \\ 
  50 & 0.954 [0.007] & 0.959 [0.006] & 0.959 [0.006] & 0.959 [0.006] \\ 
   \midrule
GAQ: 95\% CI coverage &  &  &  &  \\ 
  Studies ($k$) & $\iota^2$ = 0\% & 30\% & 60\% & 90\% \\ 
  3 & 0.999 [0.001] & 1.000 [0.000] & 0.993 [0.003] & 0.976 [0.005] \\ 
  5 & 0.997 [0.002] & 0.992 [0.003] & 0.987 [0.004] & 0.954 [0.007] \\ 
  10 & 0.972 [0.005] & 0.973 [0.005] & 0.973 [0.005] & 0.971 [0.005] \\ 
  20 & 0.972 [0.005] & 0.970 [0.005] & 0.964 [0.006] & 0.955 [0.007] \\ 
  50 & 0.960 [0.006] & 0.965 [0.006] & 0.966 [0.006] & 0.965 [0.006] \\ 
    
\multicolumn{5}{l}{\footnotesize CI = confidence interval, GAQ = global adaptive quadrature, MC = Monte Carlo.} \\ 
 \bottomrule
\end{tabular}
\end{table}

\clearpage
\clearpage
\section*{Web Appendix C: Additional Analyses of Corticosteroids and Mortality in Hospitalized COVID-19 Patients}

\begin{figure}
\centering
\begin{knitrout}
\definecolor{shadecolor}{rgb}{0.969, 0.969, 0.969}\color{fgcolor}

{\centering \includegraphics[width=0.9\linewidth]{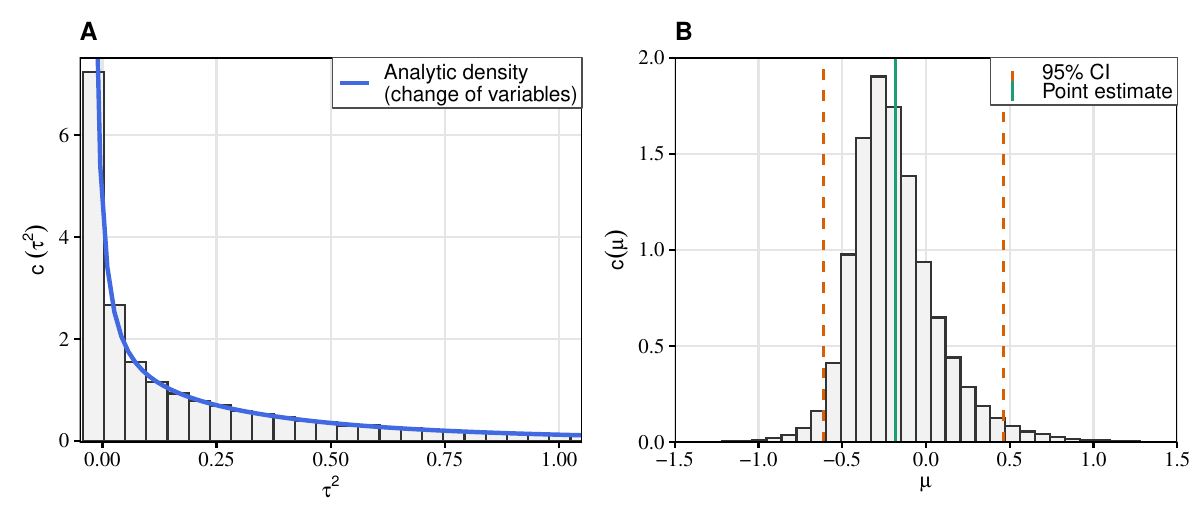} 

}

\end{knitrout}
\caption{Monte Carlo confidence distributions for the corticosteroids and COVID-19 mortality meta-analysis \citep{who2020corticosteroids}: (A) the heterogeneity parameter $\taus$, shown with its analytical confidence density derived via a change of variables (blue) and (B) the average effect $\mu$.}
\label{fig:covidconfdens}
\end{figure}

Here, we provide additional details and analyses of the case study based on seven randomized controlled trials reporting log odds ratios for the association between corticosteroids, compared with placebo or standard care, and mortality in hospitalized COVID-19 patients \citep{who2020corticosteroids}. The study-level data are summarized in Web Table~\ref{tab:coviddata}, while Web Table~\ref{tab:covidpointestimates} compares estimators of the average effect, including the CD-Edgington estimator, Edgington's estimator without heterogeneity uncertainty adjustment \citep{heldpawelhofman2024}, and the Hartung--Knapp--Sidik--Jonkman method (HKSJ; \citealp{Hartung2001, sidik2002simple}). The latter two use the Paule--Mandel heterogeneity estimator. The confidence distribution of the heterogeneity parameter $\taus$, obtained via Monte Carlo sampling together with the corresponding analytic confidence density, and the confidence distribution of the average effect $\mu$, obtained via Monte Carlo, are shown in Web Figure~\ref{fig:covidconfdens}.

Web Table~\ref{tab:serepi} reports the equi-tailed 95\% prediction intervals from the proposed PCD methods and the intervals implemented in the \texttt{R} \texttt{meta} package \citep{Skipka2006, Higgins2008, Partlett2016, Veroniki2018, Nagashima2018, Rmeta}. Edgington's PCD and the parametric bootstrap prediction intervals reflect skewness of study estimates. PCD-full and PCD-simplified intervals are wider than all comparators. This is likely because the low heterogeneity estimate ($\htaus$ = 0.03) minimally inflates the variance of future effects under $t$-distribution-based approaches, and the bootstrap method uses the exact distribution of Cochran's $\Q$ which peaks close to zero.

\begin{table}[ht]
\centering
\caption{Summary of seven randomized controlled trials on corticosteroids and mortality in hospitalized COVID-19 patients \citep{who2020corticosteroids}.} 
\label{tab:coviddata}
\begin{tabular}{lrrrr}
  \hline \\[-1em] \multicolumn{1}{l}{} & \multicolumn{2}{c} {Deaths / Patients} & \multicolumn{2}{c}{} \\ 
Study & Cx & No Cx & OR & 95\% CI \\ 
  \midrule
DEXA-COVID 19 & 2/7 & 2/12 & 2.00 & 0.21 to 18.69 \\ 
  CoDEX & 69/128 & 76/128 & 0.80 & 0.49 to 1.31 \\ 
  RECOVERY & 95/324 & 283/683 & 0.59 & 0.44 to 0.78 \\ 
  CAPE COVID & 11/75 & 20/73 & 0.46 & 0.20 to 1.04 \\ 
  COVID STEROID & 6/15 & 2/14 & 4.00 & 0.65 to 24.66 \\ 
  REMAP-CAP & 26/105 & 29/92 & 0.71 & 0.38 to 1.33 \\ 
  Steroids-SARI & 13/24 & 13/23 & 0.91 & 0.29 to 2.87 \\ 
   \multicolumn{5}{l}{\scriptsize CI = confidence interval, OR = odds ratio, Cx = Corticosteroids.} \\ 
\end{tabular}
\end{table}

\begin{table}[ht]
\centering
\caption{Point estimates and 95\% confidence intervals (CI) for the average treatment effect (log odds ratio), based on seven randomized controlled trials investigating the association between corticosteroids and mortality in hospitalized COVID-19 patients \citep{who2020corticosteroids}. Two-sided $p$-values are reported for testing $H_0: \mu = 0$ against $H_1: \mu \neq 0$.} 
\label{tab:covidpointestimates}
\begin{tabular}{lrrrrr}
  \toprule
Method & Estimate & 95\% CI & Width & Skewness & p-value \\ 
  \midrule
Hartung--Knapp--Sidik--Jonkman & -0.36 & -0.72  to  -0.00 & 0.72 & 0.00 & 0.048 \\ 
  Edgington & -0.24 & -0.54  to  0.22 & 0.76 & 0.22 & 0.22 \\ 
  CD-Edgington & -0.18 & -0.61  to  0.46 & 1.07 & 0.20 & 0.39 \\ 
  \end{tabular}
\end{table}

\begin{table}[ht]
\centering
\caption{95\% prediction intervals from the \texttt{meta} package and Edgington’s PCD distributions for seven randomized trials on corticosteroids and COVID-19 mortality \citep{who2020corticosteroids}. Medians and confidence probabilities for a future effect $\ge 0$ are also shown. Heterogeneity is estimated via Paule--Mandel \citep{paule1982consensus}, except for HTS-KR(-PR), which require REML.} 
\label{tab:serepi}
\begin{tabular}{lrrrr>{\raggedleft\arraybackslash}p{3cm}}
  \toprule
Method & Median  & 95\% PI & Width & Skewness & Conf($\tn \ge 0$) \\ 
  \midrule
PCD-full & -0.23 & -1.45 to 1.25 & 2.70 & 0.101 & 0.262 \\ 
  PCD-simplified & -0.24 & -1.50 to 1.08 & 2.58 & 0.019 & 0.245 \\ 
  PCD-fixed & -0.24 & -0.70 to 0.33 & 1.03 & 0.092 & 0.182 \\ 
  Bootstrap & -0.38 & -1.12 to 0.64 & 1.76 & 0.153 & 0.137 \\ 
  HTS-KR-PR & NC & NC &  &  &   \\ 
  HTS-KR & -0.41 & $<10^6$ to $>10^6$ & $>10^6$ & 0.000 & 0.330 \\ 
  HTS-HK-PR & -0.36 & -0.95 to 0.23 & 1.18 & 0.000 & 0.088 \\ 
  HTS-HK & -0.36 & -0.93 to 0.20 & 1.13 & 0.000 & 0.083 \\ 
  HTS-Veroniki & -0.36 & -0.93 to 0.20 & 1.13 & 0.000 & 0.083 \\ 
  HTS & -0.36 & -0.95 to 0.23 & 1.18 & 0.000 & 0.088 \\ 
  Skipka & -0.36 & -0.81 to 0.09 & 0.90 & 0.000 & 0.057 \\ 
   
\multicolumn{6}{l}{\footnotesize\makebox[0pt][l]{%
  \parbox[t]{0.9\textwidth}{HK = Hartung--Knapp, KR = Kenward--Roger, NC = non-convergence,
  PI = prediction interval, \\ PR = Partlett--Riley.}}}\\
\end{tabular}
\end{table}

\clearpage
\section*{Web Appendix D: Simulation Study Details}

\subsection*{D.1 Data-Generating Mechanism}

We vary the number of studies $k \in \{3, 5, 10, 20, 50\}$, the between-study heterogeneity determined by $\iota^2 \in \{0\%, 30\%, 60\%, 90\%\}$ \citep{higgins2025reflections}, the number of large studies $k_{\text{large}} \in \{0, 1, 2\}$, and whether true and future effects follow a normal or left-skewed skew-normal distribution. Study sizes $n_i$ are set to 50 for normal and 500 for large studies. In each iteration, we proceed as follows.

\begin{figure}
\centering
\begin{knitrout}
\definecolor{shadecolor}{rgb}{0.969, 0.969, 0.969}\color{fgcolor}

{\centering \includegraphics[width=0.5\linewidth]{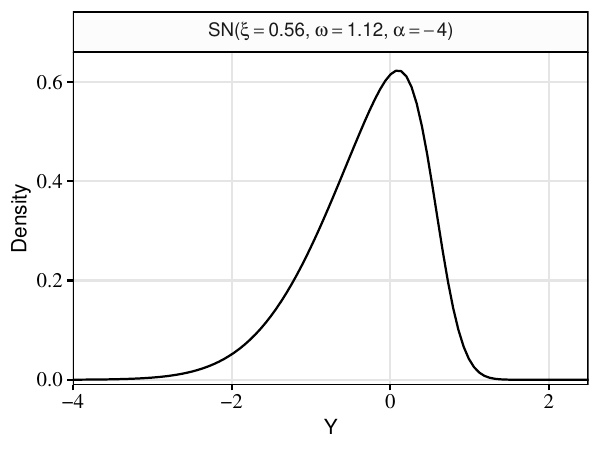} 

}

\end{knitrout}
\caption{Density of a skew-normal distribution with mean $-0.3$, variance $0.5$ and skewness parameter $\alpha = -4$.}
\label{fig:skewnormalexample}
\end{figure}

\begin{enumerate}
\item Compute $\taus$ as:
\[
\taus = \frac{1}{k} \sum_{i = 1}^k \frac{2}{n_i} \frac{\iota^2}{1 - \iota^2}.
\]
\item Simulate $k$ true effects $\ti$:
\begin{enumerate}
\item For a normal effect distribution, generate effects from a $\Nor(\mu, \taus)$.
\item For a skew-normal effect distribution, generate effects from a $\SN(\xi, \omega, \alpha)$, parameterized as by \citet{Azzalini2013}. The parameters are obtained by moment-matching such that the mean equals $-0.3$ and the variance equals $\taus$: the skewness parameter is set to $\alpha = -4$, inducing a left-skewed distribution; the scale parameter is set to $\omega = \sqrt{\taus / (1 - 2 \delta^2 / \pi)}$, where $\delta = \alpha/\sqrt{1 + \alpha^2}$; the location parameter is set to $\xi = \mu - \omega  \delta \sqrt{2 / \pi}$. An example of a skew normal distribution is displayed in Web Figure~\ref{fig:skewnormalexample}.
\end{enumerate}
\item Simulate $k$ squared standard errors $\se(\hti)^2$ from a $\Chi$-distribution:
\[
\se(\hti)^2 \sim \frac{1}{(n_i - 1)n_i} \chi_{2 (n_i - 1)}^2.
\]
\item Generate $k$ effect estimates $\hti$ on the standardized mean difference scale:
\[
\hti \sim \Nor(\ti, 2/n_i).
\]
\end{enumerate}

\subsection*{D.2 Number of Simulation Iterations}

The primary outcome of the simulation study is the coverage of 95\% prediction intervals. In each iteration \(s\), let \(\tnj\), \(j = 1, \ldots, m\), denote the generated future effects, with \(m = 10{,}000\). The estimated coverage is defined as the proportion of these future effects falling within the prediction interval \([I_\ell, I_u]\), i.e.,
\[
\coverage_s
= \frac{1}{m} \sum_{j = 1}^{m} \mathbf{1}\{ I_\ell \le \tnj \le I_u \},
\]
where \(\mathbf{1}\{\cdot\}\) denotes the indicator function. Conditional on the prediction interval, \(\coverage_s\) has variance
\[
\Var(\coverage_s) = \frac{\coverage_s (1 - \coverage_s)}{m}.
\]
This variance is maximized at \(\coverage_s = 0.5\), yielding a worst-case within-iteration standard error of \(0.005\).

Between-iteration variability was estimated from pilot simulations and reached values up to \(0.05\) for the PCD-fixed distribution with 2500 iterations. To ensure sufficient precision, we require the Monte Carlo standard error ($\semc$) of the mean coverage \(\coverage = \E[\coverage_s]\) to be at most \(0.005\). The required number of iterations is therefore estimated as
\[
\frac{\widehat{\Var}(\coverage_s)}{\semc(\coverage)^2}
= \frac{0.05}{(0.005)^2}
= 2000.
\]
To obtain stable estimates even under scenarios differing from the pilot setting, we conservatively increased the number of iterations to 4000.

\subsection*{D.3 Computational Details}
The simulation study is programmed in the \textsf{R} programming language \citep{RRR} and conducted in \textsf{R} version 4.5.0 (2025-04-11) on a remote Debian GNU/Linux server (platform: x86\_64-pc-linux-gnu). Random number generator streams are employed to ensure reproducibility in parallel execution. The code, results and detailed information on the computational environment are publicly available on Github (\url{https://github.com/davidkronthaler-dk/sim-edgemeta.git}).

\clearpage
\section*{Web Appendix E: Additional Simulation Results}

This section presents simulation results referenced but not shown in the main manuscript. Web Table~\ref{tab:ncnnf} summarizes the scenarios in which the parametric bootstrap prediction interval failed to converge at least once. Web Table~\ref{tab:appendix_summary} provides an overview of the simulation results reported in the Supplementary Material. Further, the following two sections examine the performance of the CD-Edgington estimator with respect to interval and point estimation. These results were only briefly summarized in the main manuscript and are hence described here in greater detail.

\subsection*{E.1 95\% Confidence Intervals}
\begin{landscape}
\begin{figure}
\centering
\begin{knitrout}
\definecolor{shadecolor}{rgb}{0.969, 0.969, 0.969}\color{fgcolor}

{\centering \includegraphics[width=1\linewidth]{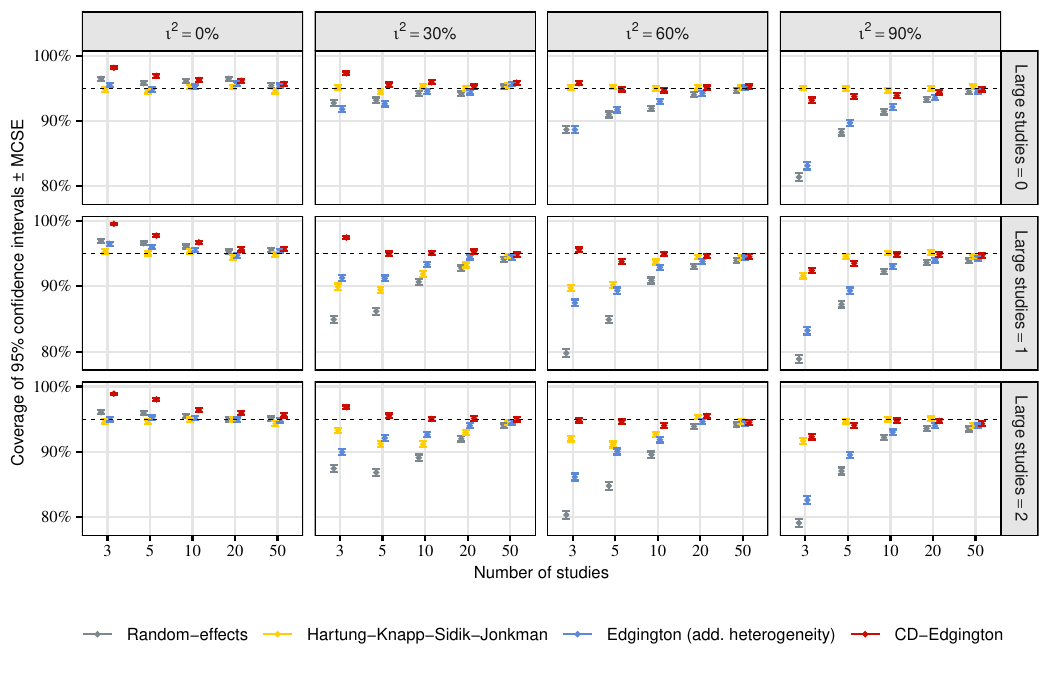} 

}

\end{knitrout}
\caption{Coverage of 95\% confidence intervals for the mean effect, for true effects following a normal distribution. Error bars represent Monte Carlo standard errors (MCSE).}
\label{fig:cicovernor}
\end{figure}
\end{landscape}

\begin{landscape}
\begin{figure}
\centering
\begin{knitrout}
\definecolor{shadecolor}{rgb}{0.969, 0.969, 0.969}\color{fgcolor}

{\centering \includegraphics[width=1\linewidth]{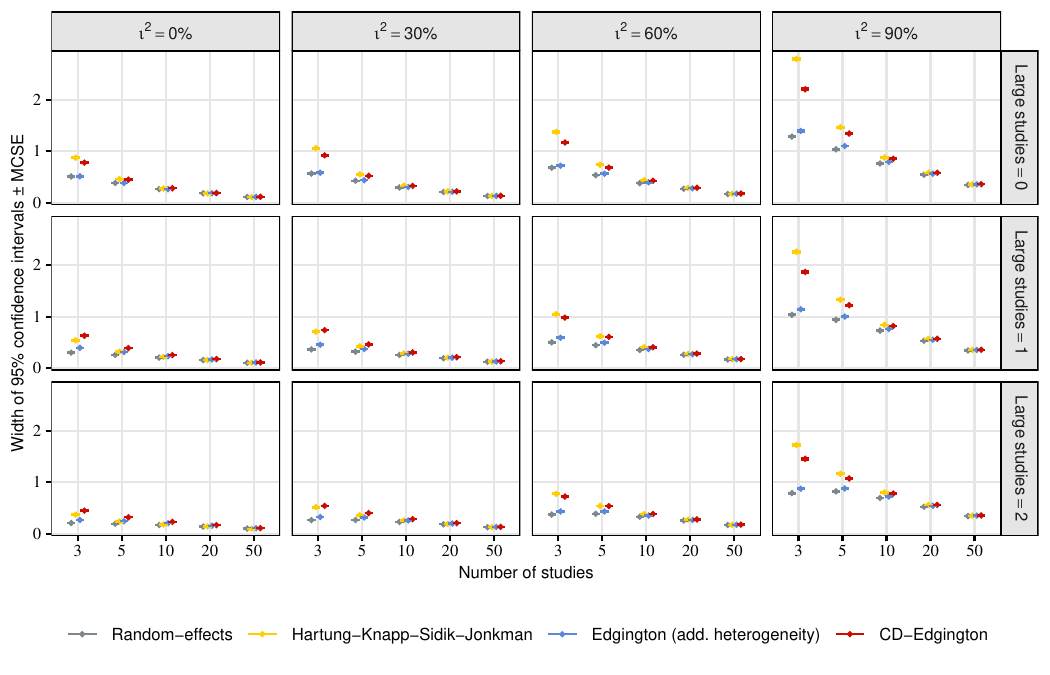} 

}

\end{knitrout}
\caption{Width of 95\% confidence intervals for the mean effect, for true effects distributed according to a normal distribution. Error bars represent Monte Carlo standard errors (MCSE).}
\label{fig:ciwnor}
\end{figure}
\end{landscape}

Web Figure~\ref{fig:cicovernor} displays the coverage of 95\% confidence intervals under normally distributed effects. Average interval widths are presented in Web Figure~\ref{fig:ciwnor}.

The CD-Edgington confidence interval tends to exhibit coverage exceeding the nominal level under no heterogeneity, approaching nominal level as the number of studies increases. The random-effects interval exhibits a similar trend but typically remains closer to nominal coverage. The Hartung--Knapp--Sidik--Jonkman (HSKJ; \citealp{Hartung2001, sidik2002simple}) method and Edgington's method with additive heterogeneity \citep{heldpawelhofman2024} typically achieve nominal coverage under no heterogeneity, with Edgington's method exceeding nominal coverage in scenarios with one large and three to five studies. Despite its conservatism, the CD-Edgington interval only marginally differs in width compared to the HKSJ interval, being narrower under no large studies and wider under one or two large studies. Edgington's method with additive heterogeneity and the random-effects interval are typically narrower when three to five studies are included, likely due to not accounting for heterogeneity estimation uncertainty. For scenarios with more than five studies, interval widths are very similar under no heterogeneity.

Under heterogeneity, the CD-Edgington interval typically attains nominal coverage, with slight overcoverage for $\iota^2$ of 30\% and three studies and slight undercoverage for $\iota^2$ of 90\% and three to five studies. The HKSJ interval attains nominal coverage in all scenarios without large studies. With one or two large studies, coverage is generally too low for three to ten studies, except under $\iota^2$ of 90\%, where undercoverage occurs only with three studies.

Edgington's method with additive heterogeneity adjustment typically approaches nominal coverage under heterogeneity provided that at least ten studies are included. For fewer studies, coverage may be as low as 85\%. In scenarios with large studies and heterogeneity it typically outperforms the random-effects interval but yields consistently lower coverage than the CD-Edgington interval, and typically also lower coverage than the HKSJ method. The random-effects interval exhibits substantial undercoverage with ten or fewer studies under heterogeneity.

The CD-Edgington interval is typically narrower than the HKSJ interval, particularly evident under large heterogeneity, except in scenarios with three studies, one or two large studies and  $\iota^2$ of 0\% or 30\%. Both methods produce confidence intervals that are typically wider than the approaches not accounting for uncertainty in heterogeneity estimation (i.e., the random-effects interval and Edgington's method with additive heterogeneity), with differences diminishing as the number of studies increases.

\begin{landscape}
\begin{figure}
\centering
\begin{knitrout}
\definecolor{shadecolor}{rgb}{0.969, 0.969, 0.969}\color{fgcolor}

{\centering \includegraphics[width=1\linewidth]{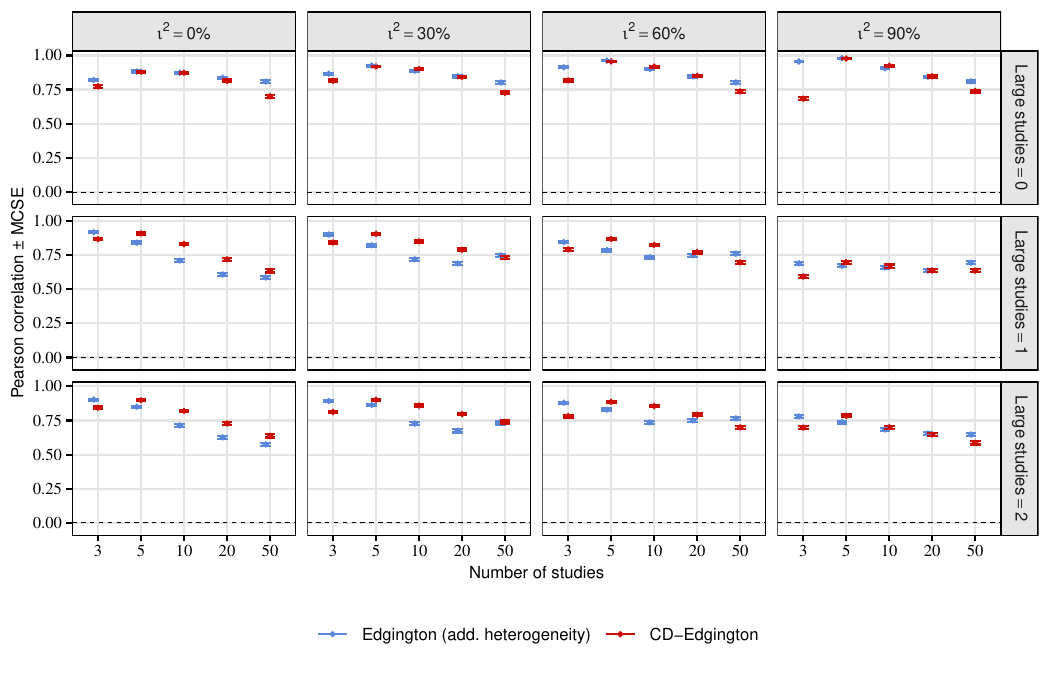} 

}

\end{knitrout}
\caption{Pearson correlation between the skewness of 95\% confidence intervals and the skewness of effect estimates, for true effects distributed according to a normal distribution. Error bars represent Monte Carlo standard errors (MCSE).}
\label{fig:ciskhesnor}
\end{figure}
\end{landscape}

\begin{landscape}
\begin{figure}
\centering
\begin{knitrout}
\definecolor{shadecolor}{rgb}{0.969, 0.969, 0.969}\color{fgcolor}

{\centering \includegraphics[width=1\linewidth]{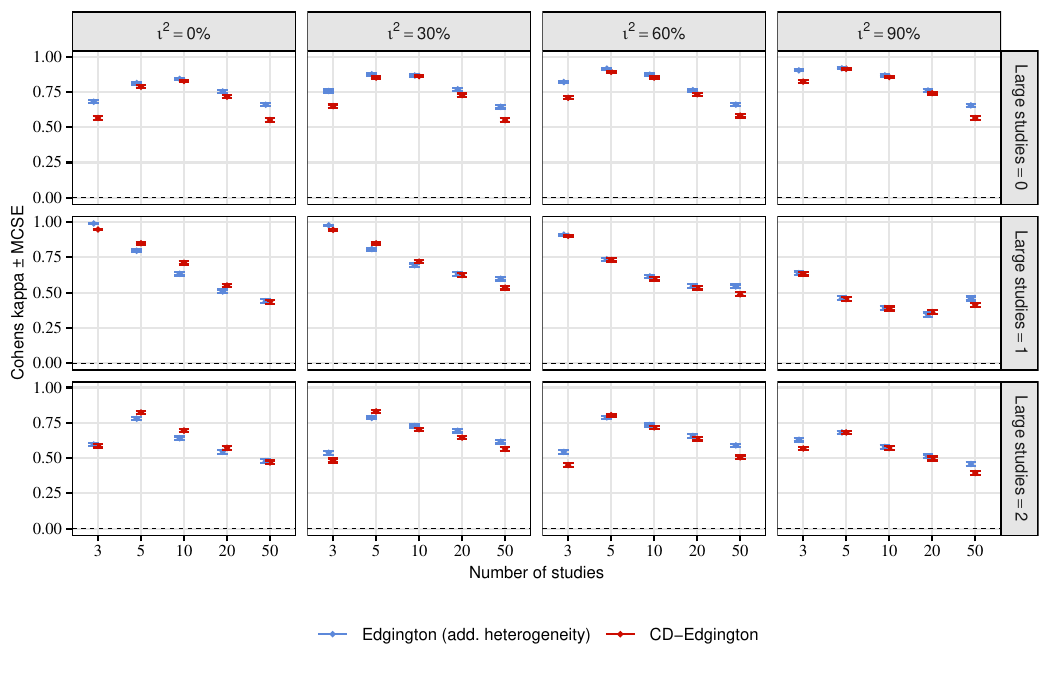} 

}

\end{knitrout}
\caption{Cohens kappa for sign agreement between the skewness of 95\% confidence intervals and the skewness of effect estimates, for true effects distributed according to a normal distribution. Error bars represent Monte Carlo standard errors (MCSE).}
\label{fig:cikappanor}
\end{figure}
\end{landscape}

Web Figure~\ref{fig:ciskhesnor} displays the Pearson correlation between the skewness of confidence intervals and effect estimates for normally distributed effects. Note that the the random-effects interval and the HKSJ interval are not included since they are always symmetric. Both approaches based on Edgington's method reflect the skewness of effect estimates effectively, with correlations never falling below 0.5 and sometimes approaching one. Correlations generally decrease as the number of studies increases. Under three or 50 studies, Edgington's method with additive heterogeneity typically exhibits larger correlations, while the CD-Edington interval does so for five to 20 studies. Similar trends are observed for Cohen's kappa assessing sign agreement (Web Figure~\ref{fig:cikappanor}).

\begin{landscape}
\begin{figure}
\centering
\begin{knitrout}
\definecolor{shadecolor}{rgb}{0.969, 0.969, 0.969}\color{fgcolor}

{\centering \includegraphics[width=1\linewidth]{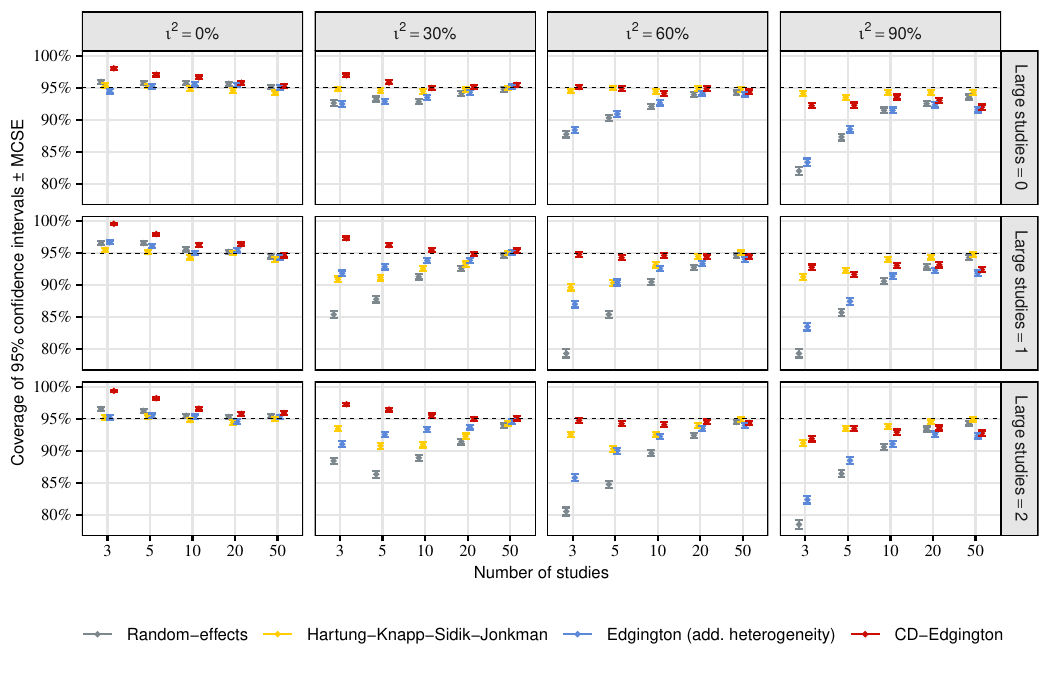} 

}

\end{knitrout}
\caption{Coverage of 95\% confidence intervals for the mean effect, for true effects following a left-skewed skew-normal distribution. Error bars represent Monte Carlo standard errors (MCSE).}
\label{fig:cicoversn}
\end{figure}
\end{landscape}

\begin{landscape}
\begin{figure}
\centering
\begin{knitrout}
\definecolor{shadecolor}{rgb}{0.969, 0.969, 0.969}\color{fgcolor}

{\centering \includegraphics[width=1\linewidth]{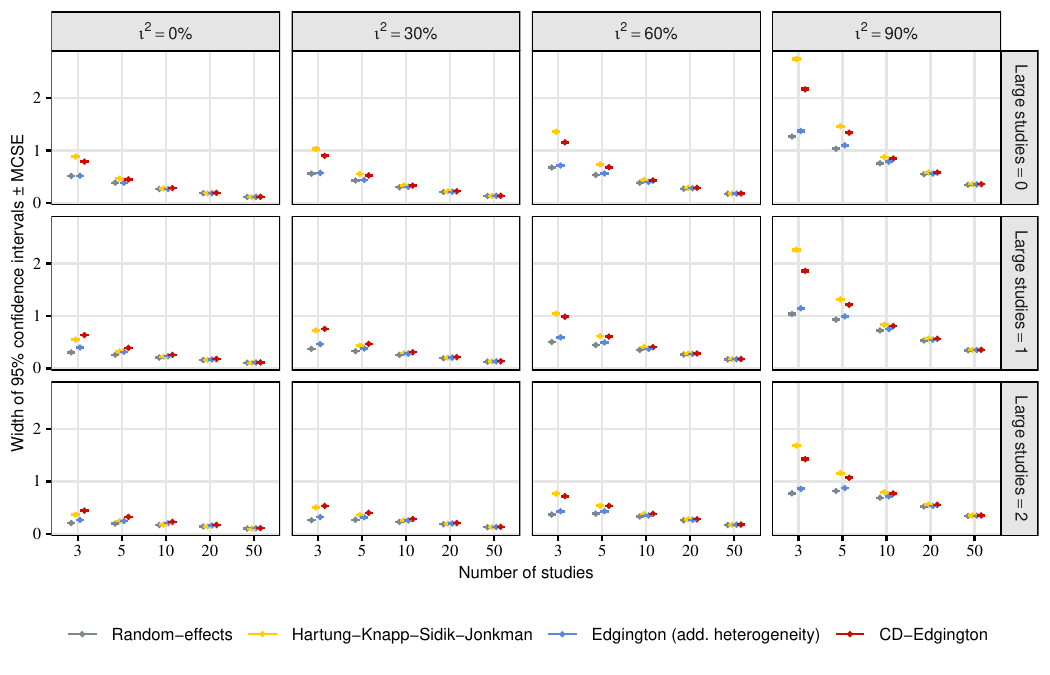} 

}

\end{knitrout}
\caption{Width of 95\% confidence intervals for the mean effect, for true effects distributed according to a left-skewed skew-normal distribution. Error bars represent Monte Carlo standard errors (MCSE).}
\label{fig:ciwsn}
\end{figure}
\end{landscape}

\begin{landscape}
\begin{figure}
\centering
\begin{knitrout}
\definecolor{shadecolor}{rgb}{0.969, 0.969, 0.969}\color{fgcolor}

{\centering \includegraphics[width=1\linewidth]{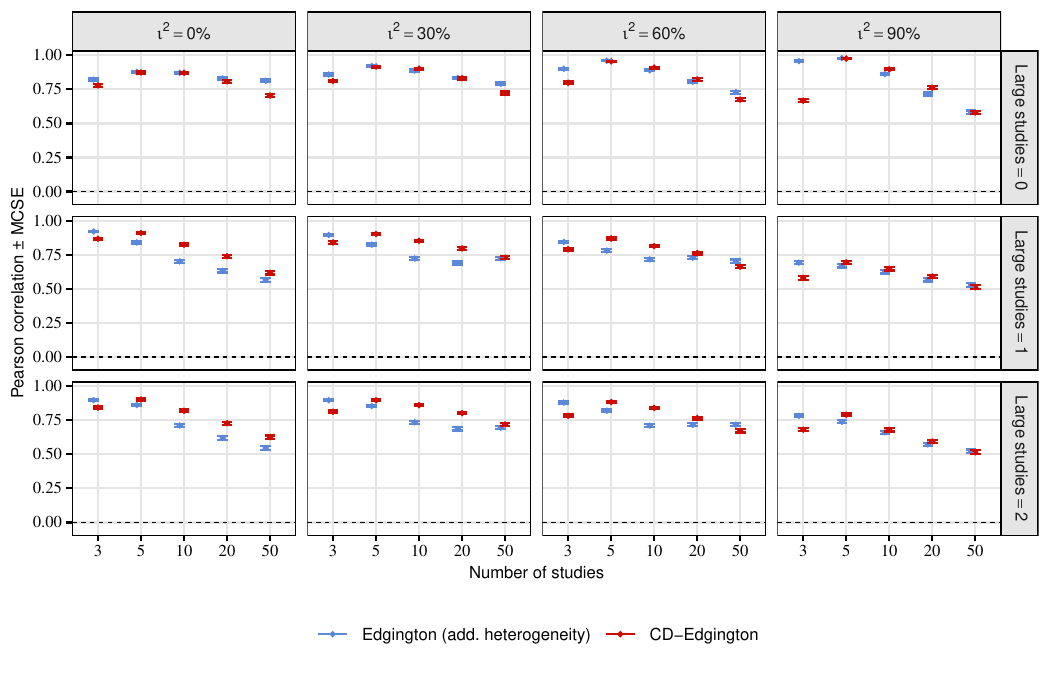} 

}

\end{knitrout}
\caption{Pearson correlation between the skewness of 95\% confidence intervals and the skewness of effect estimates, for true effects distributed according to a left-skewed skew-normal distribution. Error bars represent Monte Carlo standard errors (MCSE).}
\label{fig:ciskhessn}
\end{figure}
\end{landscape}

\begin{landscape}
\begin{figure}
\centering
\begin{knitrout}
\definecolor{shadecolor}{rgb}{0.969, 0.969, 0.969}\color{fgcolor}

{\centering \includegraphics[width=1\linewidth]{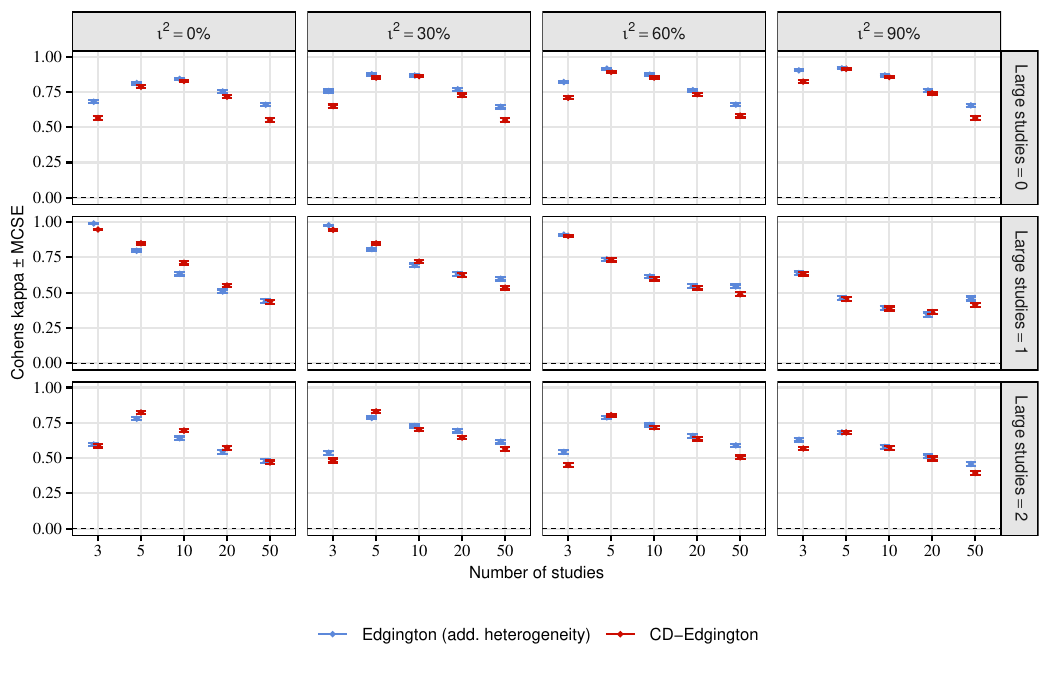} 

}

\end{knitrout}
\caption{Cohens kappa for sign agreement between the skewness of 95\% confidence intervals and the skewness of effect estimates, for true effects distributed according to a left-skewed skew-normal distribution. Error bars represent Monte Carlo standard errors (MCSE).}
\label{fig:cikappasn}
\end{figure}
\end{landscape}

For true effects distributed according to a skew-normal distribution, Web Figures~\ref{fig:cicoversn},~\ref{fig:ciwsn},~\ref{fig:ciskhessn} and~\ref{fig:cikappasn} display corresponding results. Coverages are similar across effect distributions, with only slight decreases for effects distributed according to a skew-normal distribution, mainly observed under  $\iota^2$ of 90\%. Confidence interval widths and skewness results are virtually identical under both effect distributions.

\subsection*{E.2 Point Estimation}
\begin{landscape}
\begin{figure}
\centering
\begin{knitrout}
\definecolor{shadecolor}{rgb}{0.969, 0.969, 0.969}\color{fgcolor}

{\centering \includegraphics[width=1\linewidth]{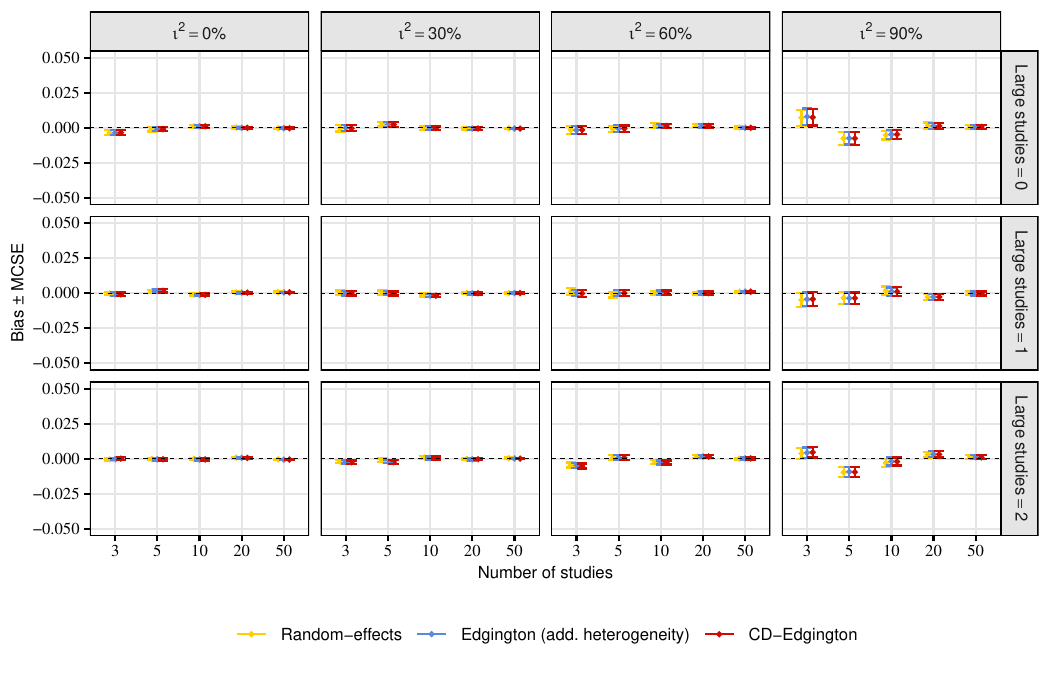} 

}

\end{knitrout}
\caption{Bias for the mean effect, for true effects distributed according to a normal distribution. Error bars represent Monte Carlo standard errors (MCSE).}
\label{fig:biasnor}
\end{figure}
\end{landscape}

\begin{landscape}
\begin{figure}
\centering
\begin{knitrout}
\definecolor{shadecolor}{rgb}{0.969, 0.969, 0.969}\color{fgcolor}

{\centering \includegraphics[width=1\linewidth]{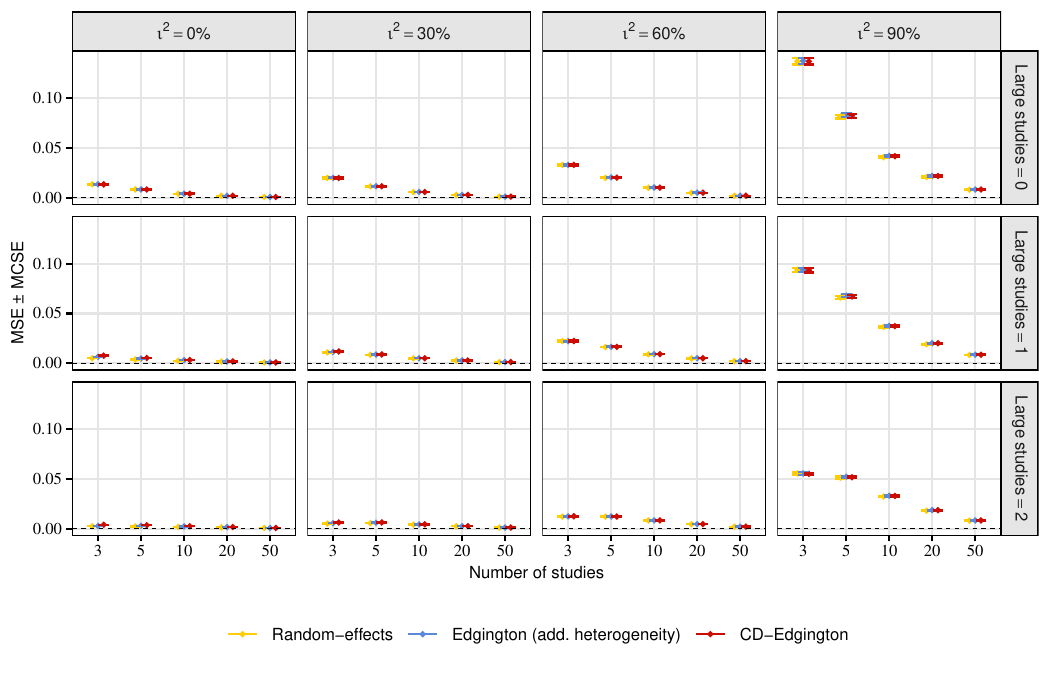} 

}

\end{knitrout}
\caption{Mean squared error (MSE) for the mean effect, for true effects distributed according to a normal distribution. Error bars represent Monte Carlo standard errors (MCSE).}
\label{fig:msenor}
\end{figure}
\end{landscape}

The average bias of point estimators under normally distributed effects is presented in Web Figure~\ref{fig:biasnor}. All estimators (Monte Carlo CD-Edgington, classical inverse-variance weights estimator, Edgington's method with additive heterogeneity; \citealp{heldpawelhofman2024}) are approximately unbiased for the true mean effect. Fewer studies and larger heterogeneity increase variability, while increasing the number of studies generally reduces bias. The largest average bias observed occurs under  $\iota^2$ of 90\%, reaching $-0.0097$ for the random-effects estimator. The corresponding MSEs are comparable across methods and decrease as the number of studies increases and increase with larger heterogeneity (Web Figure~\ref{fig:msenor}).

\begin{landscape}
\begin{figure}
\centering
\begin{knitrout}
\definecolor{shadecolor}{rgb}{0.969, 0.969, 0.969}\color{fgcolor}

{\centering \includegraphics[width=1\linewidth]{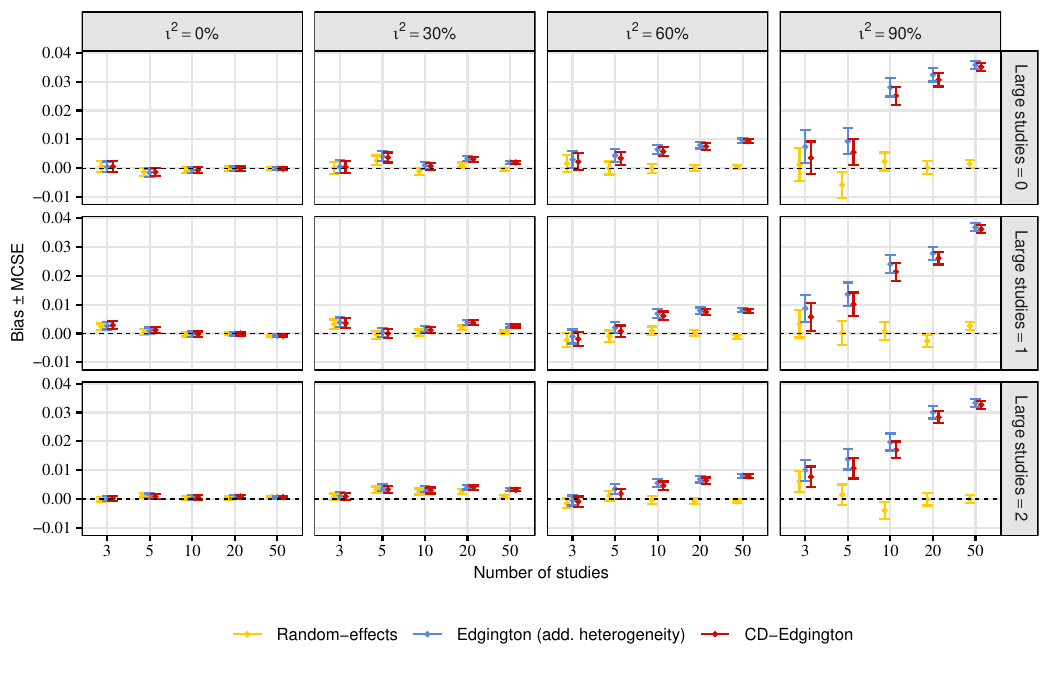} 

}

\end{knitrout}
\caption{Bias for the mean effect, for true effects distributed according to a left-skewed skew-normal distribution. Error bars represent Monte Carlo standard errors (MCSE).}
\label{fig:biassn}
\end{figure}
\end{landscape}

When true effects follow a skew-normal distribution, the bias of methods based on Edgington's approach systematically deviates from zero under scenarios with  $\iota^2$ of 60\% and 90\% (Web Figure~\ref{fig:biassn}). Notably, the bias increases in the number of studies when $\iota^2$ is 90\%. Maximum average bias of 0.037 is observed for Edgington's method with additive heterogeneity adjustment under $\iota^2$ of 90\% and 50 studies. The average bias is consistently positive in these scenarios, reflecting the left-skewness of the skew-normal distribution. If the skew-normal distribution were right-skewed, we would expect the bias to be negative instead. In contrast, the random-effects estimator remains approximately unbiased across all scenarios. MSE trends resemble those observed under the normal effect distribution, with slightly higher MSEs under a skew-normal effect distribution (Web Figure~\ref{fig:msenor}).\\

\begin{table}[ht]
\centering
\caption{Scenarios of the simulation study under which the 95\% parametric bootstrap prediction interval did not converge at least once.} 
\label{tab:ncnnf}
\begin{tabular}{rrrlr}
  \toprule
Studies & $\iota^2$ (\%) & Large studies & Distribution & Non-convergences \\ 
  \midrule
3 & 30 & 0 & Skew-normal &   1 \\ 
  3 & 30 & 1 & Normal &   1 \\ 
  3 & 60 & 0 & Normal &   1 \\ 
  3 & 60 & 1 & Skew-normal &   3 \\ 
  3 & 60 & 2 & Normal &   2 \\ 
  3 & 60 & 2 & Skew-normal &   1 \\ 
  3 & 90 & 0 & Normal &   1 \\ 
  3 & 90 & 0 & Skew-normal &   9 \\ 
  3 & 90 & 1 & Normal &   4 \\ 
  3 & 90 & 1 & Skew-normal &   7 \\ 
  3 & 90 & 2 & Normal &   4 \\ 
  3 & 90 & 2 & Skew-normal &   3 \\ 
  \end{tabular}
\end{table}

\renewcommand{\arraystretch}{1.3}

\newcommand{\sidehead}[2]{%
  \cellcolor{gray!20}%
  \adjustbox{angle=90,lap=0pt}{%
    \parbox[c][#1][c]{1cm}{\centering\bfseries #2}%
  }%
}

\begin{table}[ht]
\centering
\caption{Simulation results presented in the Supplementary Material.}
\normalsize
\begin{tabularx}{\linewidth}{@{}p{1.4cm} X r r@{}}
\hline
& \textbf{Performance measure} & \textbf{Effect distribution} & \textbf{Display} \\
\hline

\multirow{10}{*}{\vertlabel{10cm}{Prediction}}
& Coverage of 95\% prediction intervals
& Skew-normal
& \ref{fig:picoversn} \\
& Coverage distributions of 95\% prediction intervals ($\iota^2$ of 0\%)
& No effect distribution
& \ref{fig:coverdist_nor0} \\
& Coverage distributions of 95\% prediction intervals ($\iota^2$ of 30\%)
& Normal / Skew-normal
& \ref{fig:coverdist_nor30} / \ref{fig:coverdist_sn30} \\
& Coverage distributions of 95\% prediction intervals ($\iota^2$ of 60\%)
& Normal / Skew-normal
& \ref{fig:coverdist_nor60} / \ref{fig:coverdist_sn60} \\
& Coverage distributions of 95\% prediction intervals ($\iota^2$ of 90\%)
& Normal / Skew-normal
& \ref{fig:coverdist_nor90} / \ref{fig:coverdist_sn90} \\
& Pearson correlation between skewness of 95\% prediction intervals and skewness of effect estimates
& Skew-normal
& \ref{fig:piskewhessn} \\
& Cohen's kappa for sign agreement between skewness of 95\% prediction intervals and skewness of effect estimates
& Normal / Skew-normal
& \ref{fig:pikappanor} / \ref{fig:pikappasn} \\
& Width of 95\% prediction intervals
& Normal / Skew-normal
& \ref{fig:piwnor} / \ref{fig:piwsn} \\
& Continuous ranked probability scores
& Normal / Skew-normal
& \ref{fig:crpsnor} / \ref{fig:crpssn} \\
& Computation time
& Normal / Skew-normal
& \ref{fig:comptimenor} / \ref{fig:comptimesn} \\
\hline

\multirow{7}{*}{\vertlabel{5.55cm}{Estimation}}
& Coverage of 95\% confidence intervals
& Normal / Skew-normal
& \ref{fig:cicovernor} / \ref{fig:cicoversn} \\
& Width of 95\% confidence intervals
& Normal / Skew-normal
& \ref{fig:ciwnor} / \ref{fig:ciwsn} \\
& Pearson correlation between skewness of 95\% confidence intervals and skewness of effect estimates
& Normal / Skew-normal
& \ref{fig:ciskhesnor} / \ref{fig:ciskhessn} \\
& Cohen's kappa for sign agreement between skewness of 95\% confidence intervals and skewness of effect estimates
& Normal / Skew-normal
& \ref{fig:cikappanor} / \ref{fig:cikappasn} \\
& Bias of point estimators
& Normal / Skew-normal
& \ref{fig:biasnor} / \ref{fig:biassn} \\
& Mean squared error (MSE) of point estimators
& Normal / Skew-normal
& \ref{fig:msenor} / \ref{fig:msesn} \\
& Computation time 
& No effect distribution
& \ref{tab:timeCDEdgington} \\
\hline
\end{tabularx}
\label{tab:appendix_summary}
\end{table}

\begin{landscape}
\begin{figure}
\centering
\begin{knitrout}
\definecolor{shadecolor}{rgb}{0.969, 0.969, 0.969}\color{fgcolor}

{\centering \includegraphics[width=1\linewidth]{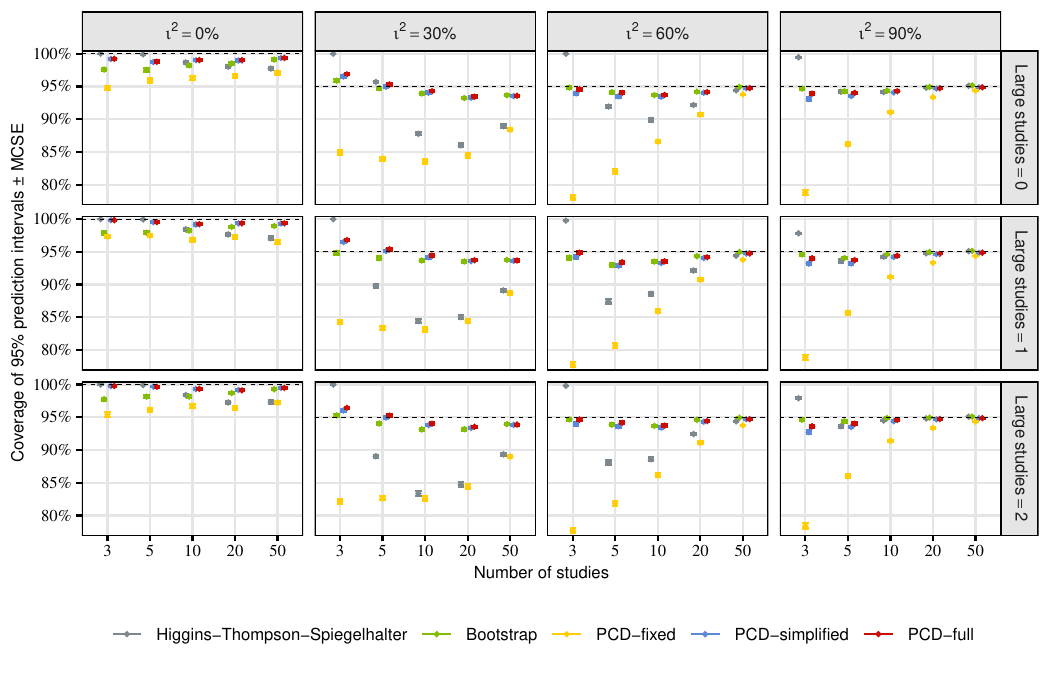} 

}

\end{knitrout}
\caption{Coverage of 95\% prediction intervals for true effects following a left-skewed skew-normal distribution. Error bars represent Monte Carlo standard errors (MCSE). For no heterogeneity ($\iota^2 = 0\%$), future effects coincide with the common effect, with 100\% coverage reflecting that prediction intervals cover the common effect.}
\label{fig:picoversn}
\end{figure}
\end{landscape}

\begin{landscape}
\begin{figure}
\centering
\begin{knitrout}
\definecolor{shadecolor}{rgb}{0.969, 0.969, 0.969}\color{fgcolor}

{\centering \includegraphics[width=1\linewidth]{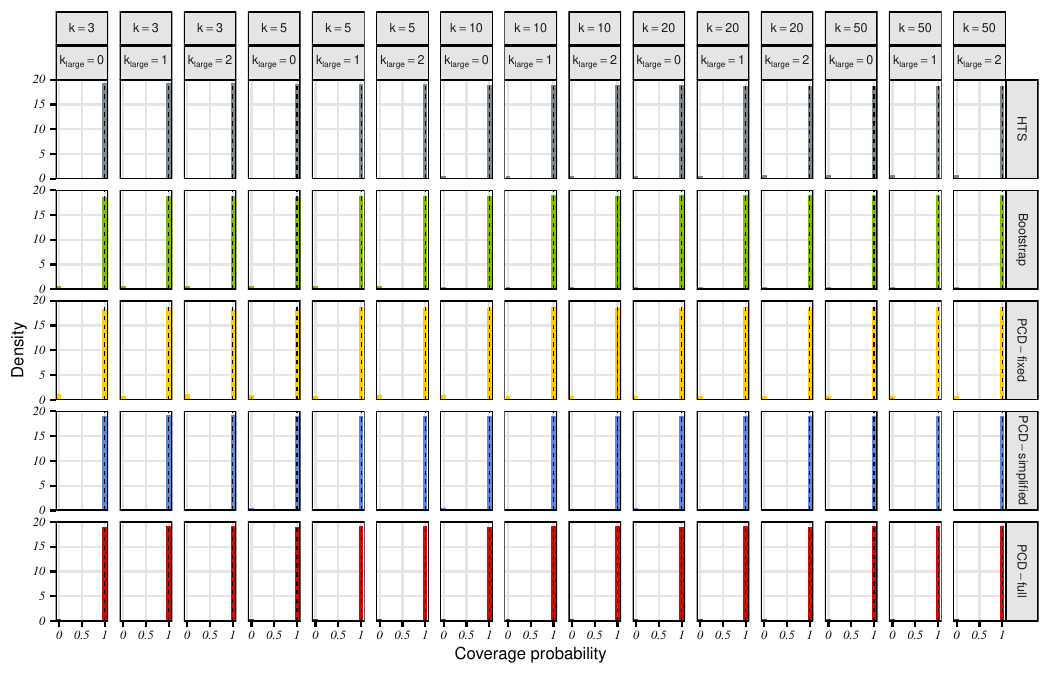} 

}

\end{knitrout}
\caption{Distribution of coverage probabilities of 95\% prediction intervals when  $\iota^2$ is 0\%. For no heterogeneity, future effects coincide with the common effect, with 100\% coverage (vertical dashed line) reflecting that prediction intervals cover the common effect. Histograms are displayed on the density scale (20 bins over the interval $[0,1]$), so densities can reach values around 20 when observed coverage probabilities concentrate strongly near 1.}

\label{fig:coverdist_nor0}
\end{figure}
\end{landscape}

\begin{landscape}
\begin{figure}
\centering
\begin{knitrout}
\definecolor{shadecolor}{rgb}{0.969, 0.969, 0.969}\color{fgcolor}

{\centering \includegraphics[width=1\linewidth]{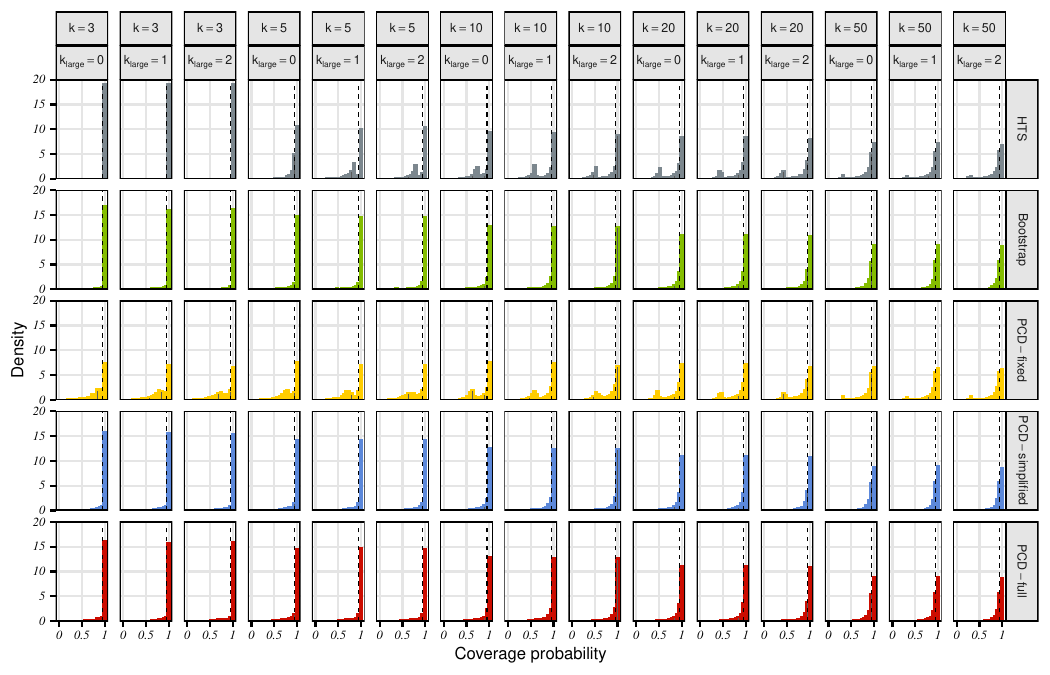} 

}

\end{knitrout}
\caption{Distribution of coverage probabilities of 95\% prediction intervals, for true effects following a normal distribution and  $\iota^2$ of 30\%. Histograms are displayed on the density scale (20 bins over the interval $[0,1]$), so densities can reach values around 20 when observed coverage probabilities concentrate strongly near 1. Vertical dashed lines indicate the nominal coverage probability of 0.95.} 
\label{fig:coverdist_nor30}
\end{figure}
\end{landscape}

\begin{landscape}
\begin{figure}
\centering
\begin{knitrout}
\definecolor{shadecolor}{rgb}{0.969, 0.969, 0.969}\color{fgcolor}

{\centering \includegraphics[width=1\linewidth]{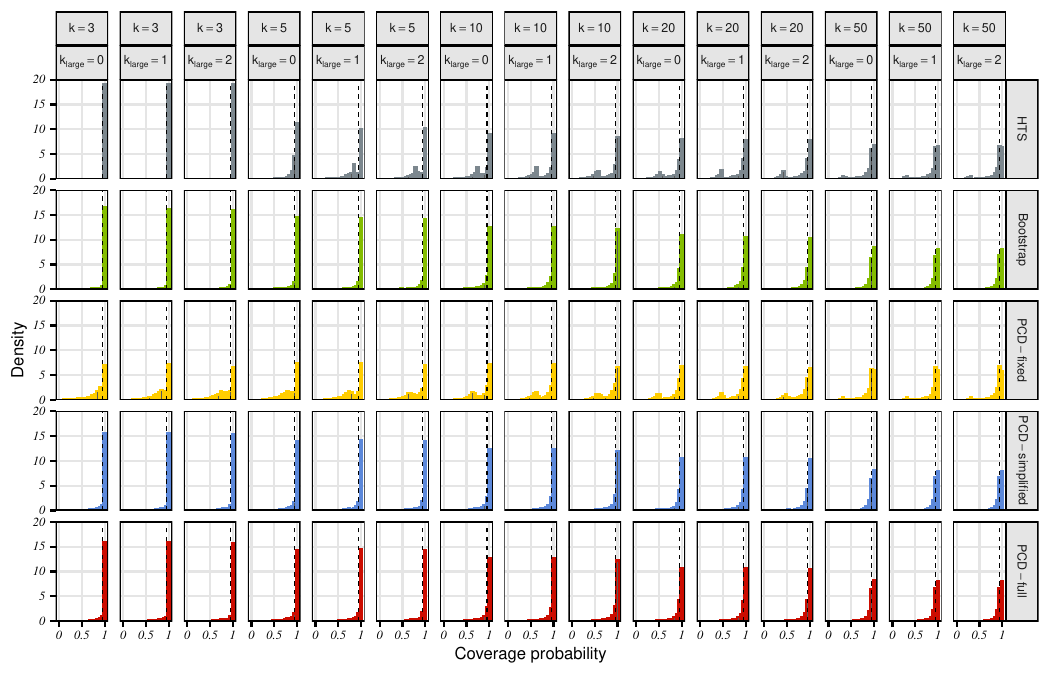} 

}

\end{knitrout}
\caption{Distribution of coverage probabilities of 95\% prediction intervals, for true effects following a left-skewed skew-normal distribution and  $\iota^2$ of 30\%. Histograms are displayed on the density scale (20 bins over the interval $[0,1]$), so densities can reach values around 20 when observed coverage probabilities concentrate strongly near 1. Vertical dashed lines indicate the nominal coverage probability of 0.95.} 
\label{fig:coverdist_sn30}
\end{figure}
\end{landscape}

\begin{landscape}
\begin{figure}
\centering
\begin{knitrout}
\definecolor{shadecolor}{rgb}{0.969, 0.969, 0.969}\color{fgcolor}

{\centering \includegraphics[width=1\linewidth]{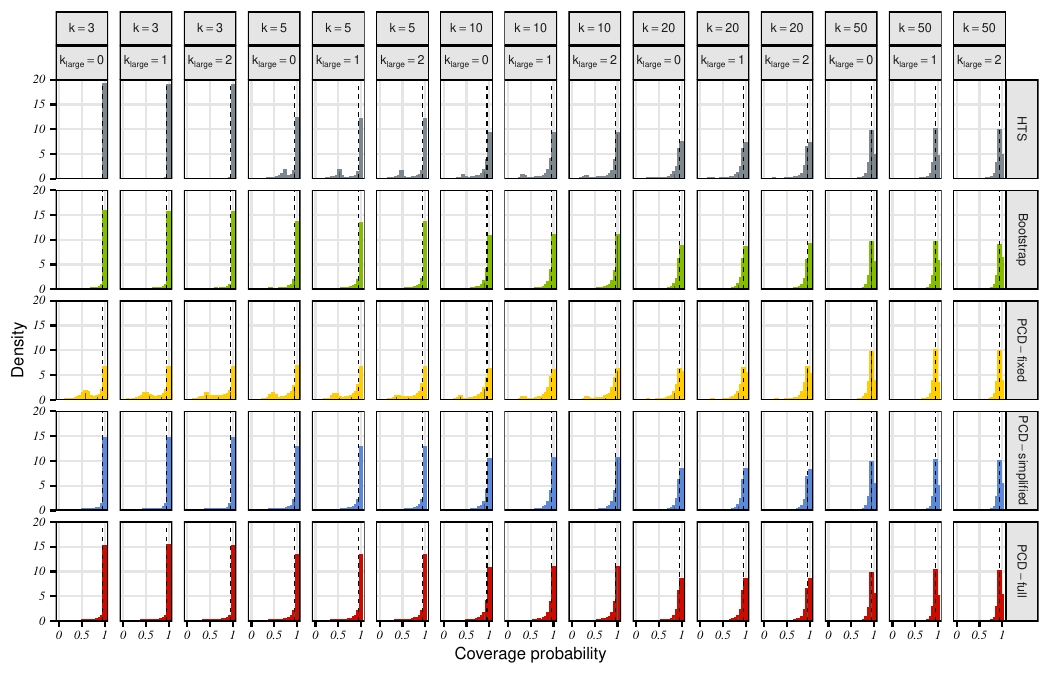} 

}

\end{knitrout}
\caption{Distribution of coverage probabilities of 95\% prediction intervals, for true effects following a normal distribution and  $\iota^2$ of 60\%. Histograms are displayed on the density scale (20 bins over the interval $[0,1]$), so densities can reach values around 20 when observed coverage probabilities concentrate strongly near 1. Vertical dashed lines indicate the nominal coverage probability of 0.95.} 
\label{fig:coverdist_nor60}
\end{figure}
\end{landscape}

\begin{landscape}
\begin{figure}
\centering
\begin{knitrout}
\definecolor{shadecolor}{rgb}{0.969, 0.969, 0.969}\color{fgcolor}

{\centering \includegraphics[width=1\linewidth]{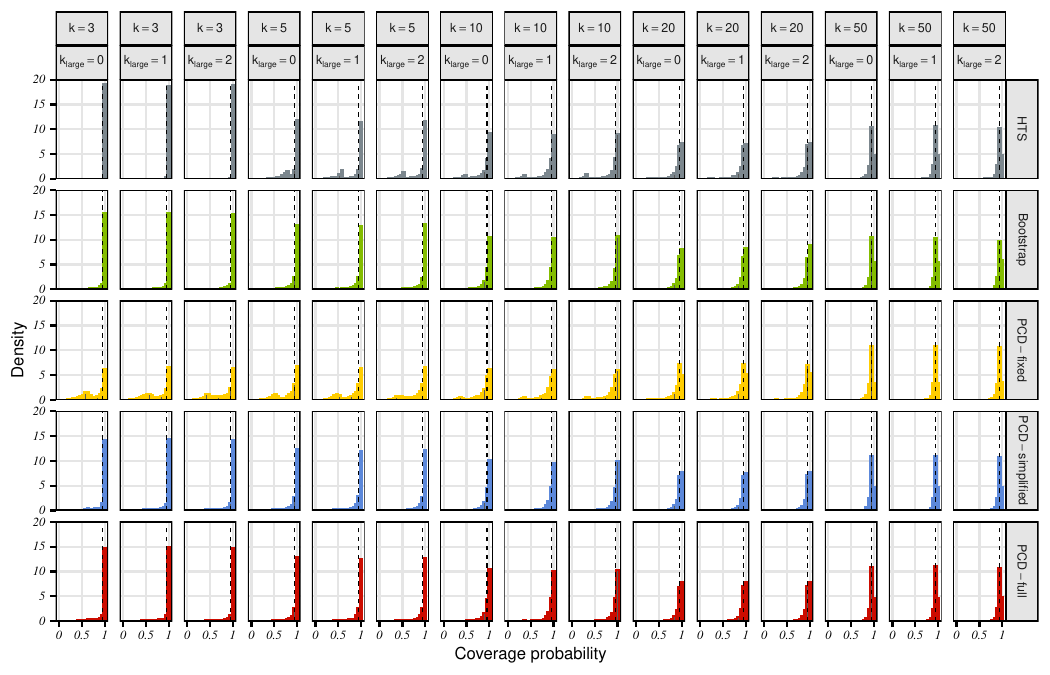} 

}

\end{knitrout}
\caption{Distribution of coverage probabilities of 95\% prediction intervals, for true effects following a left-skewed skew-normal distribution and  $\iota^2$ of 60\%. Histograms are displayed on the density scale (20 bins over the interval $[0,1]$), so densities can reach values around 20 when observed coverage probabilities concentrate strongly near 1. Vertical dashed lines indicate the nominal coverage probability of 0.95.} 
\label{fig:coverdist_sn60}
\end{figure}
\end{landscape}

\begin{landscape}
\begin{figure}
\centering
\begin{knitrout}
\definecolor{shadecolor}{rgb}{0.969, 0.969, 0.969}\color{fgcolor}

{\centering \includegraphics[width=1\linewidth]{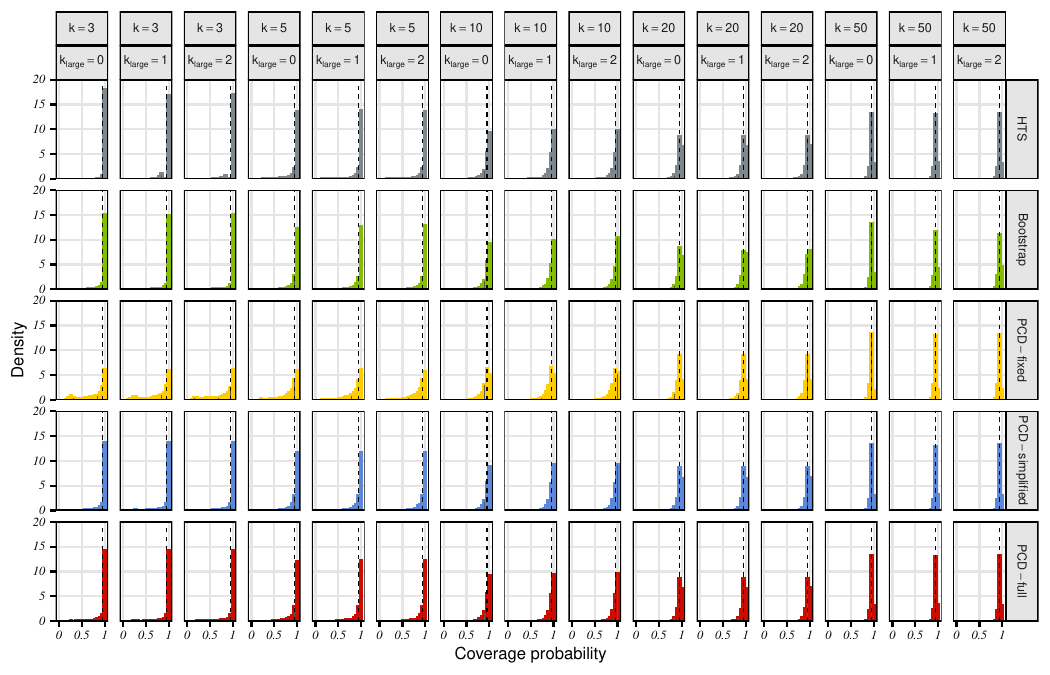} 

}

\end{knitrout}
\caption{Distribution of coverage probabilities of 95\% prediction intervals, for true effects following a normal distribution and  $\iota^2$ of 90\%. Histograms are displayed on the density scale (20 bins over the interval $[0,1]$), so densities can reach values around 20 when observed coverage probabilities concentrate strongly near 1. Vertical dashed lines indicate the nominal coverage probability of 0.95.} 
\label{fig:coverdist_nor90}
\end{figure}
\end{landscape}

\begin{landscape}
\begin{figure}
\centering
\begin{knitrout}
\definecolor{shadecolor}{rgb}{0.969, 0.969, 0.969}\color{fgcolor}

{\centering \includegraphics[width=1\linewidth]{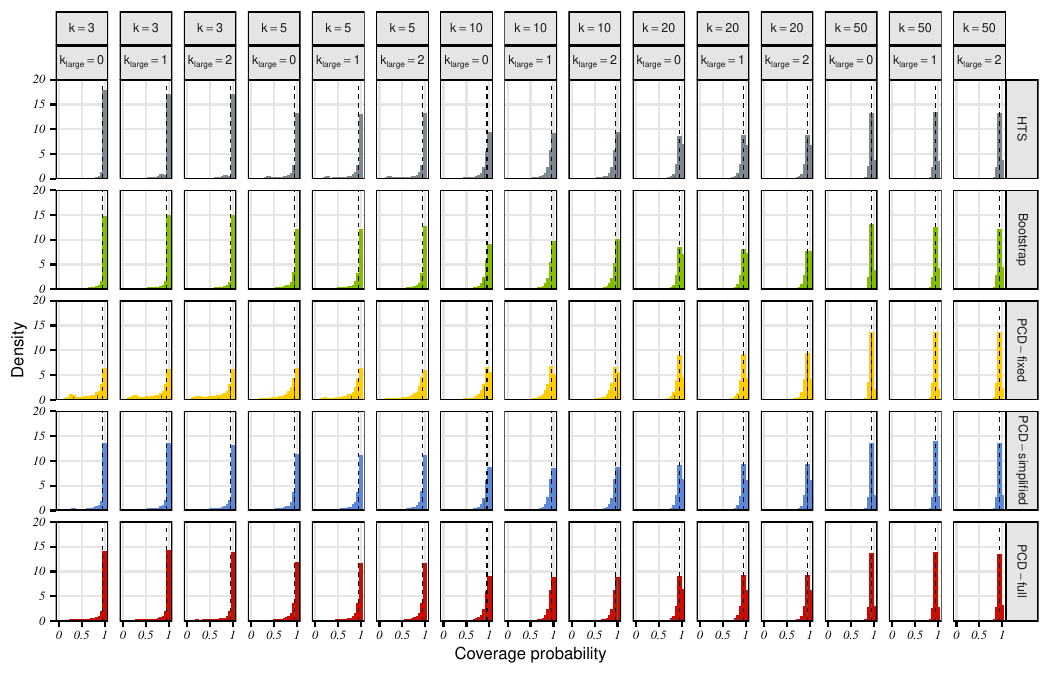} 

}

\end{knitrout}
\caption{Distribution of coverage probabilities of 95\% prediction intervals, for true effects following a left-skewed skew-normal distribution and  $\iota^2$ of 90\%. Histograms are displayed on the density scale (20 bins over the interval $[0,1]$), so densities can reach values around 20 when observed coverage probabilities concentrate strongly near 1. Vertical dashed lines indicate the nominal coverage probability of 0.95.} 
\label{fig:coverdist_sn90}
\end{figure}
\end{landscape}

\begin{landscape}
\begin{figure}
\centering
\begin{knitrout}
\definecolor{shadecolor}{rgb}{0.969, 0.969, 0.969}\color{fgcolor}

{\centering \includegraphics[width=1\linewidth]{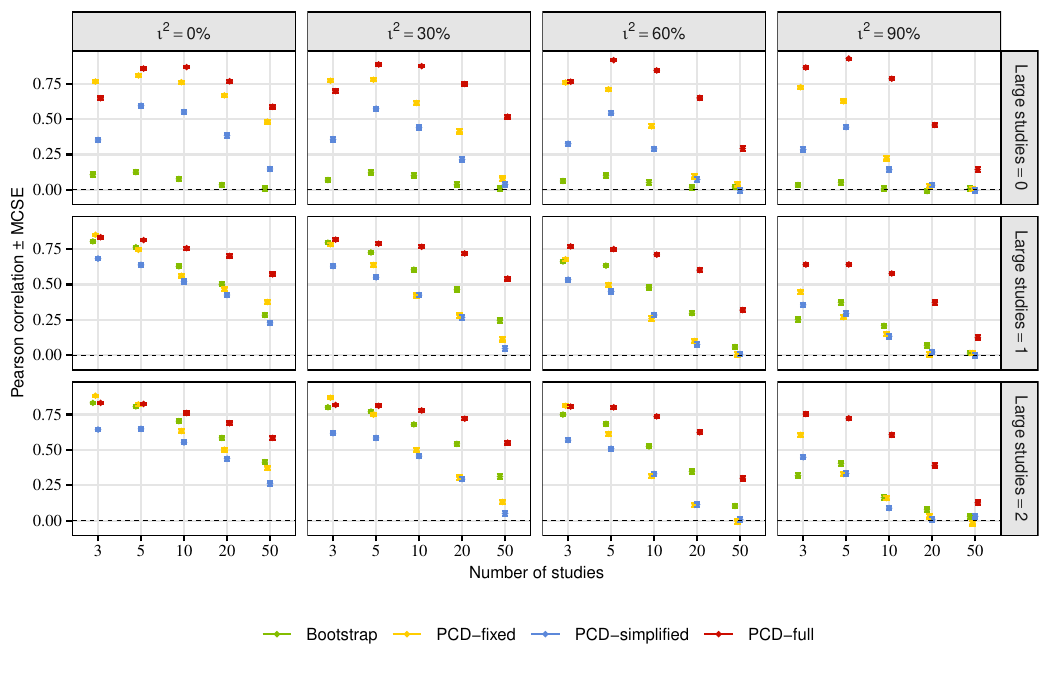} 

}

\end{knitrout}
\caption{Pearson correlation between the skewness of 95\% prediction intervals and the skewness of effect estimates for true effects following a left-skewed skew-normal distribution. Error bars represent Monte Carlo standard errors (MCSE).}
\label{fig:piskewhessn}
\end{figure}
\end{landscape}

\begin{landscape}
\begin{figure}
\centering
\begin{knitrout}
\definecolor{shadecolor}{rgb}{0.969, 0.969, 0.969}\color{fgcolor}

{\centering \includegraphics[width=1\linewidth]{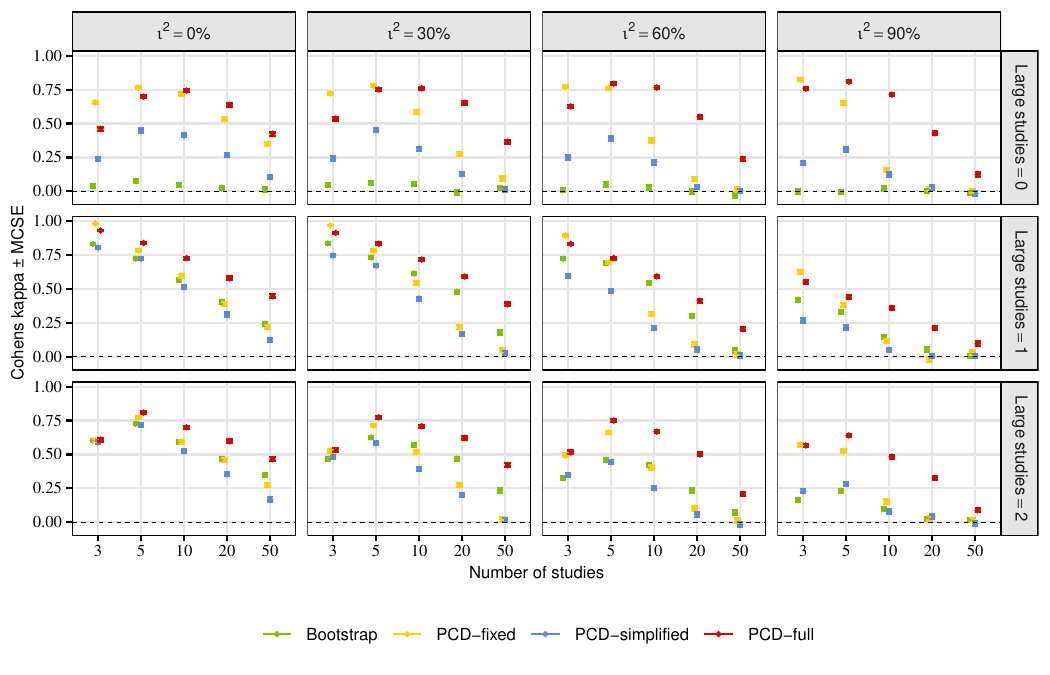} 

}

\end{knitrout}
\caption{Cohens kappa quantifying sign agreement between the skewness of 95\% prediction intervals and the skewness of effect estimates, for true effects distributed according to a normal distribution. Error bars represent Monte Carlo standard errors (MCSE).}
\label{fig:pikappanor}
\end{figure}
\end{landscape}

\begin{landscape}
\begin{figure}
\centering
\begin{knitrout}
\definecolor{shadecolor}{rgb}{0.969, 0.969, 0.969}\color{fgcolor}

{\centering \includegraphics[width=1\linewidth]{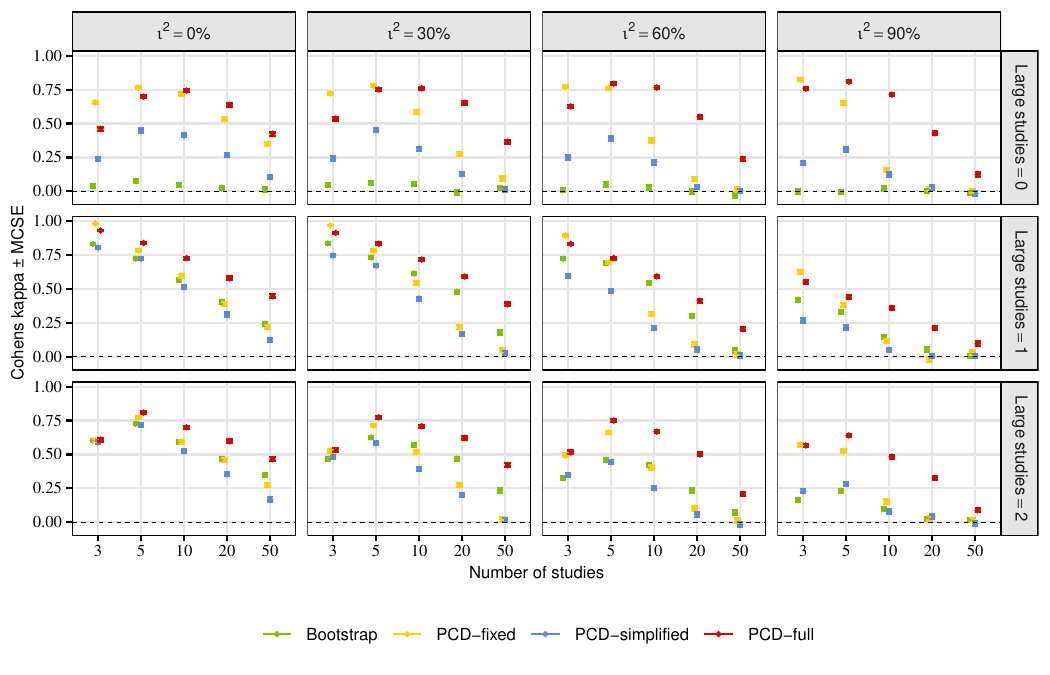} 

}

\end{knitrout}
\caption{Cohens kappa quantifying sign agreement between the skewness of 95\% prediction intervals and the skewness of effect estimates, for true effects distributed according to a skew-normal distribution. Error bars represent Monte Carlo standard errors (MCSE).}
\label{fig:pikappasn}
\end{figure}
\end{landscape}

\begin{landscape}
\begin{figure}
\centering
\begin{knitrout}
\definecolor{shadecolor}{rgb}{0.969, 0.969, 0.969}\color{fgcolor}

{\centering \includegraphics[width=1\linewidth]{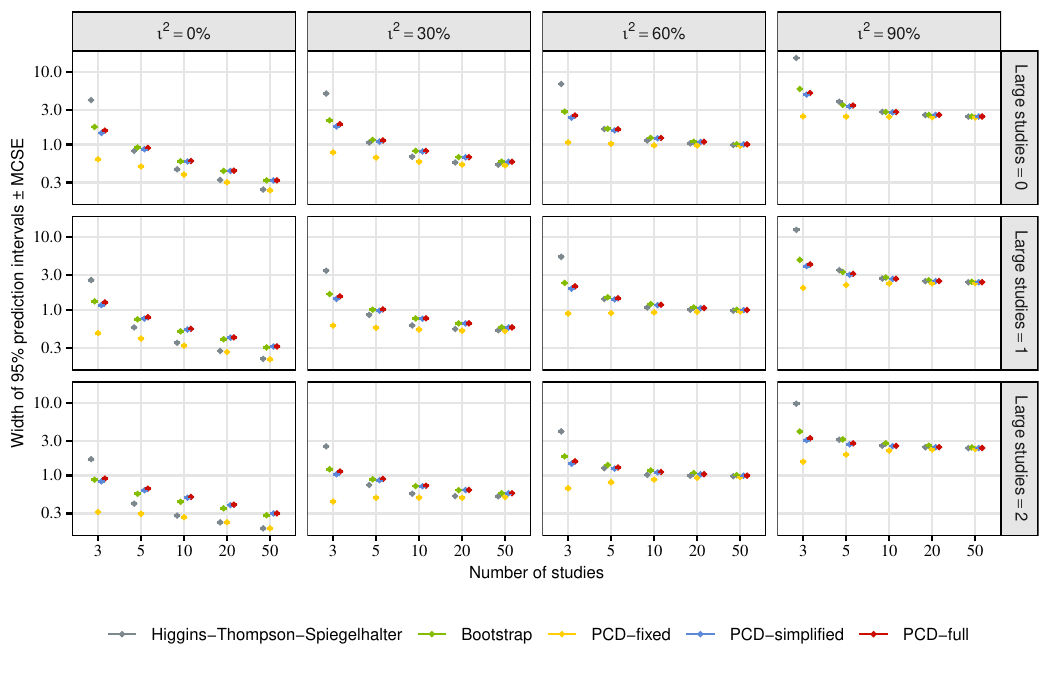} 

}

\end{knitrout}
\caption{Width of 95\% prediction intervals for true effects following a normal distribution. Error bars represent Monte Carlo standard errors (MCSE). The y-axis is displayed spaced on a log$_{10}$ scale (without transformation of the data) to facilitate comparison.}
\label{fig:piwnor}
\end{figure}
\end{landscape}

\begin{landscape}
\begin{figure}
\centering
\begin{knitrout}
\definecolor{shadecolor}{rgb}{0.969, 0.969, 0.969}\color{fgcolor}

{\centering \includegraphics[width=1\linewidth]{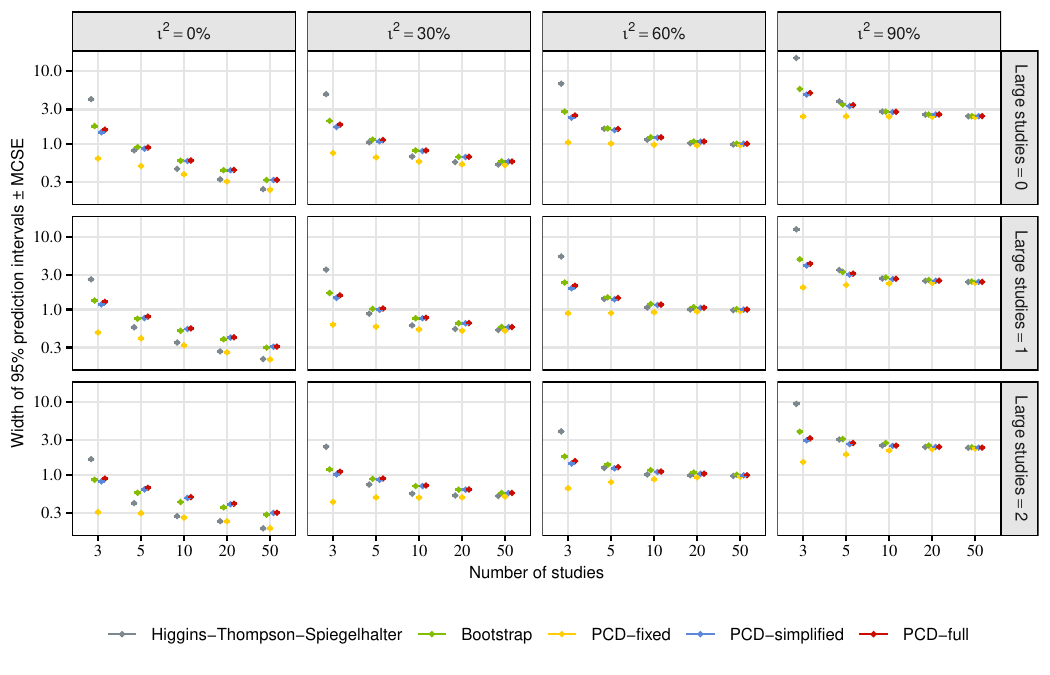} 

}

\end{knitrout}
\caption{Width of 95\% prediction intervals for true effects following a left-skewed skew-normal distribution. Error bars represent Monte Carlo standard errors (MCSE). The y-axis is displayed spaced on a log$_{10}$ scale (without transformation of the data) to facilitate comparison.}
\label{fig:piwsn}
\end{figure}
\end{landscape}

\begin{landscape}
\begin{figure}
\centering
\begin{knitrout}
\definecolor{shadecolor}{rgb}{0.969, 0.969, 0.969}\color{fgcolor}

{\centering \includegraphics[width=1\linewidth]{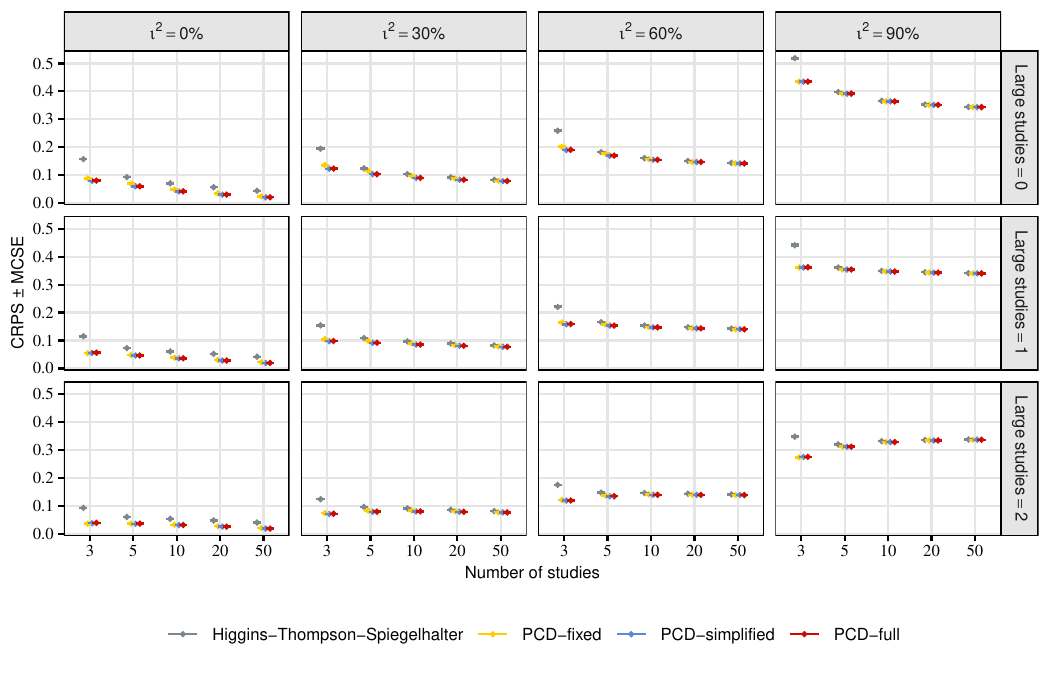} 

}

\end{knitrout}
\caption{Continous ranked probability scores (CRPS) for true effects following a normal distribution. Error bars represent Monte Carlo standard errors (MCSE).}
\label{fig:crpsnor}
\end{figure}
\end{landscape}

\begin{landscape}
\begin{figure}
\centering
\begin{knitrout}
\definecolor{shadecolor}{rgb}{0.969, 0.969, 0.969}\color{fgcolor}

{\centering \includegraphics[width=1\linewidth]{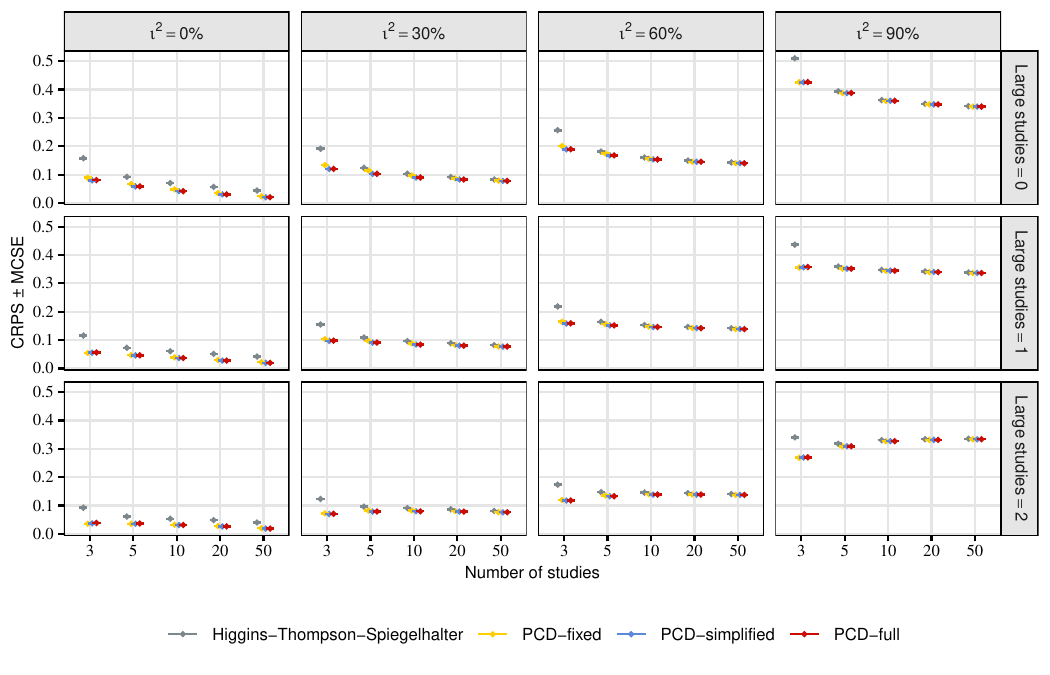} 

}

\end{knitrout}
\caption{Continous ranked probability scores (CRPS) for true effects following a left-skewed skew-normal distribution. Error bars represent Monte Carlo standard errors (MCSE).}
\label{fig:crpssn}
\end{figure}
\end{landscape}

\begin{landscape}
\begin{figure}
\centering
\begin{knitrout}
\definecolor{shadecolor}{rgb}{0.969, 0.969, 0.969}\color{fgcolor}

{\centering \includegraphics[width=1\linewidth]{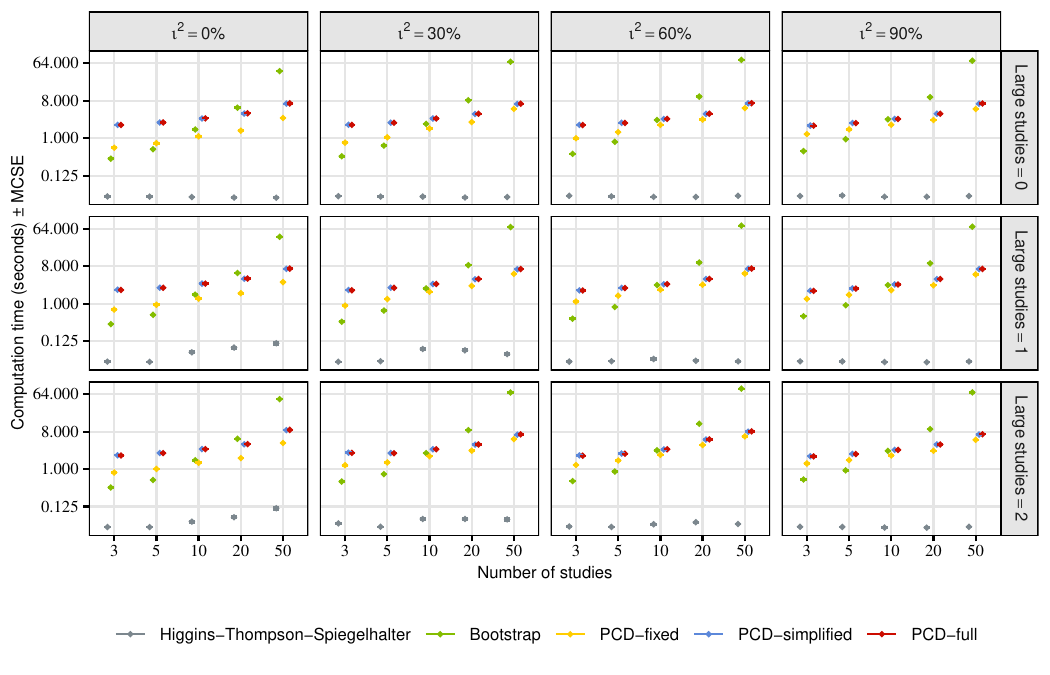} 

}

\end{knitrout}
\caption{Mean computation time for 95\% prediction intervals for true effects following a normal distribution. Error bars represent Monte Carlo standard errors (MCSE). The y-axis is displayed spaced on a log$_{2}$ scale (without transformation of the data) to facilitate comparison.}
\label{fig:comptimenor}
\end{figure}
\end{landscape}

\begin{landscape}
\begin{figure}
\centering
\begin{knitrout}
\definecolor{shadecolor}{rgb}{0.969, 0.969, 0.969}\color{fgcolor}

{\centering \includegraphics[width=1\linewidth]{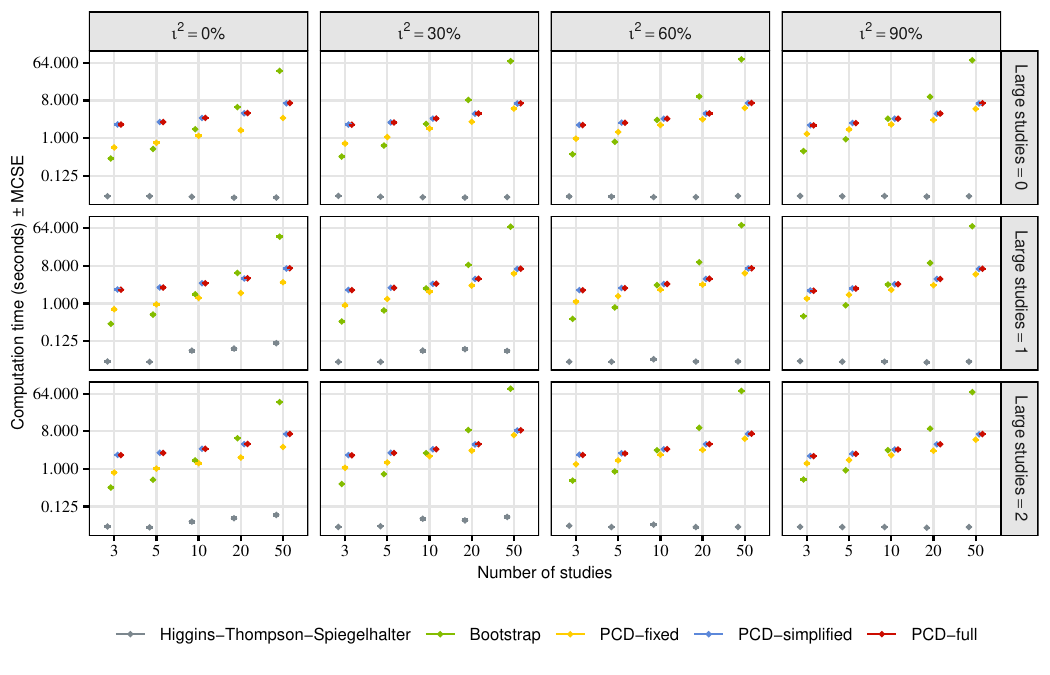} 

}

\end{knitrout}
\caption{Mean computation time for 95\% prediction intervals for true effects following a skew-normal distribution. Error bars represent Monte Carlo standard errors (MCSE). The y-axis is displayed spaced on a log$_{2}$ scale (without transformation of the data) to facilitate comparison.}
\label{fig:comptimesn}
\end{figure}
\end{landscape}

\begin{landscape}
\begin{figure}
\centering
\begin{knitrout}
\definecolor{shadecolor}{rgb}{0.969, 0.969, 0.969}\color{fgcolor}

{\centering \includegraphics[width=1\linewidth]{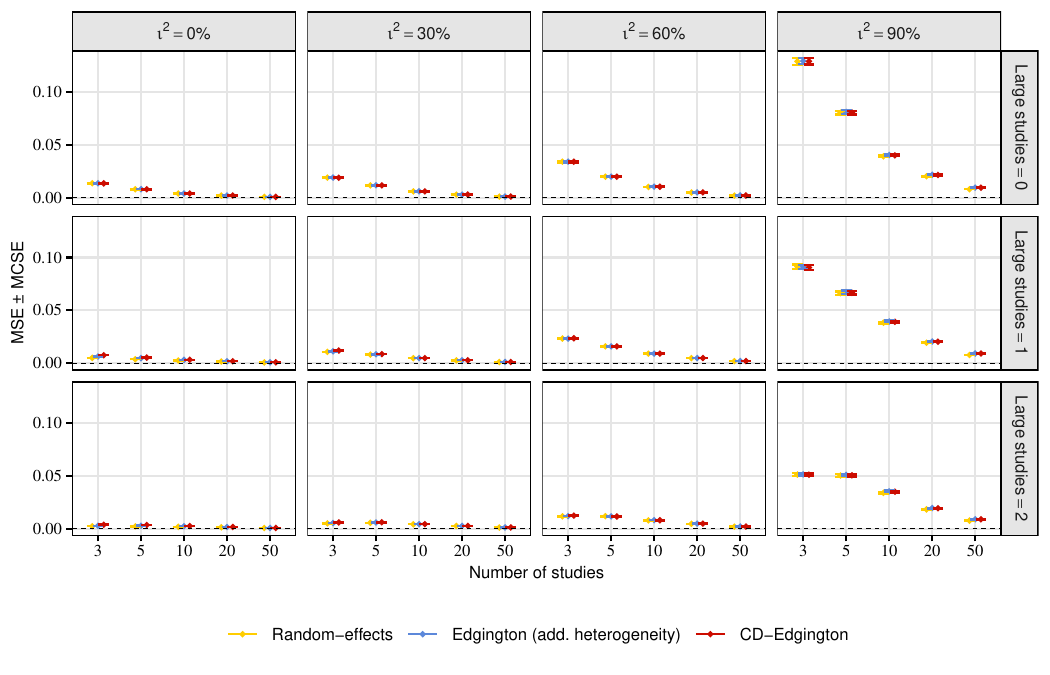} 

}

\end{knitrout}
\caption{Mean squared error (MSE) for the mean effect, for true effects distributed according to a left-skewed skew-normal distribution. Error bars represent Monte Carlo standard errors (MCSE).}
\label{fig:msesn}
\end{figure}
\end{landscape}

\begin{table}[ht]
\centering
\caption{Mean computation time (seconds; s) with Monte Carlo standard error (MCSE) of the Monte Carlo CD-Edgington estimator across varying numbers of studies, based on 1000 simulation iterations.} 
\label{tab:timeCDEdgington}
\begin{tabular}{lr}
  \toprule
Studies ($k$) & Runtime (s) [MCSE] \\ 
  \midrule
3 & 2.93 [0.02] \\ 
  5 & 3.37 [0.03] \\ 
  10 & 4.54 [0.03] \\ 
  20 & 6.20 [0.04] \\ 
  50 & 10.94 [0.07] \\ 
  \end{tabular}
\end{table}

\clearpage
\section*{Web Appendix F: Weighted Edgington Methods}
\subsection*{F.1 Weighted Edgington Confidence and Predictive Distributions}
The methods for computing predictive distributions and the CD-Edgington estimator can be extended to incorporate study-specific weights by replacing the unweighted combined $p$-value function (Equation~\eqref{eq:pEdgington} in the main text) with its weighted analogue, following \citet{BarrowSmith1979}. For weights $w_i, i \in \{1,\ldots,k\}$, the weighted Edgington combined $p$-value function (Pawel et al., manuscript in preparation)
 is given by 
\begin{equation*}
p_w(\mu) =
\begin{cases}
\displaystyle
\frac{1}{k!\prod_{i=1}^k w_i}
\sum_{\mathbf v \in \{0,1\}^k}
(-1)^{\sum_{i=1}^k v_i}
\left\{ s_w(\mu) - \mathbf w^\top \mathbf v \right\}_+^k
& \text{if } k_{\mathrm{eff}} < 12, \\[1.2em]
\displaystyle
\Phi\left(
\frac{
s_w(\mu) - \frac{1}{2}\sum_{i=1}^k w_i
}{
\sqrt{\frac{1}{12}\sum_{i=1}^k w_i^2}
}
\right)
& \text{if } k_{\mathrm{eff}} \ge 12,
\end{cases}
\end{equation*}
where $\{x\}_+ := \max(x,0)$ denotes the positive-part operator,
\[
s_w(\mu) = \sum_{i=1}^k w_i \, p_{1\text{s},+,i}(\mu)
\qquad \text{and} \qquad
k_{\mathrm{eff}}
=
\frac{\left(\sum_{i=1}^k w_i^2\right)^2}{\sum_{i=1}^k w_i^4}.
\]
The study-specific $p$-value functions $p_{1\text{s},+,i}(\mu)$ are computed as in Equation~\eqref{eq:marginal} of the main text, while the weighted combined $p$-value function replaces its unweighted counterpart in subsequent steps. In particular, step~1 of the Monte Carlo sampling algorithm (Algorithm~\ref{alg:mc_sampling} in the main text), namely sampling from the confidence distribution of the heterogeneity parameter, remains unchanged. In step~2, the weighted combined $p$-value function defines the confidence distribution of the average effect, from which samples are generated via the probability integral transform. Predictive samples are then obtained as before. For the CD-Edgington estimator computed using GAQ integration, the weighted combined $p$-value function similarly defines the confidence distribution function, whose derivative yields the corresponding confidence density used in the marginalization step.

\subsection*{F.2 Weighted Analysis of the COVID-19 Corticosteroid Meta-Analysis}

\begin{figure}
\centering
\begin{knitrout}
\definecolor{shadecolor}{rgb}{0.969, 0.969, 0.969}\color{fgcolor}

{\centering \includegraphics[width=1\linewidth]{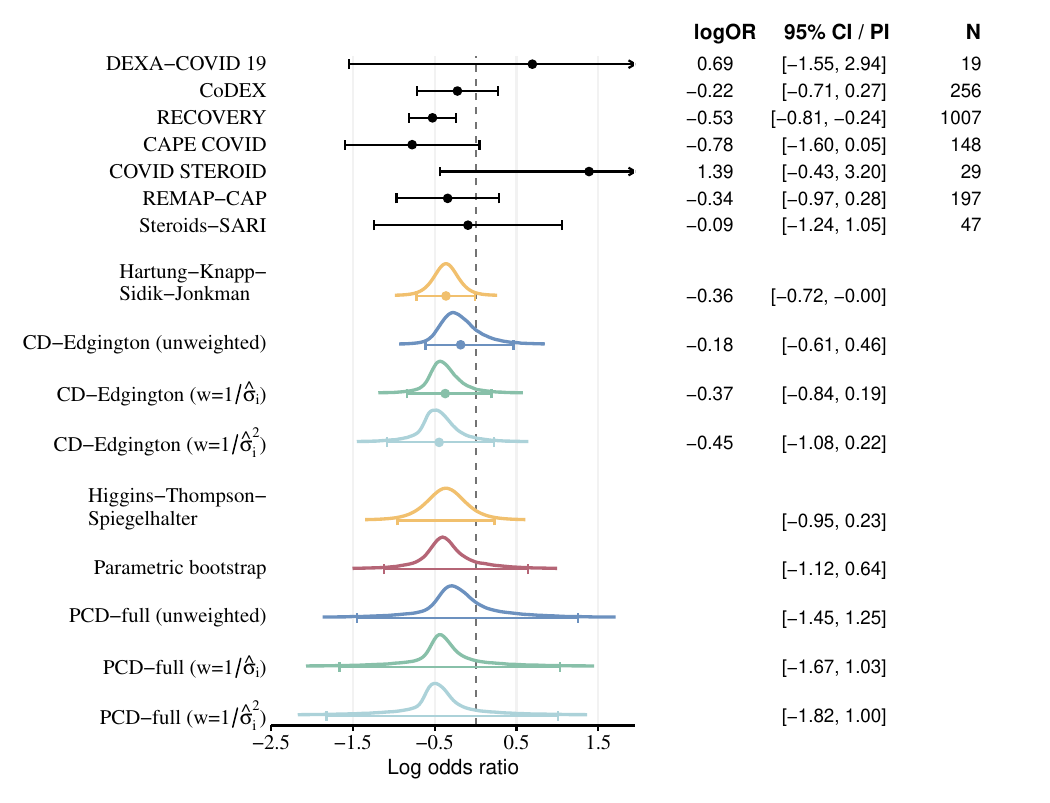} 

}

\end{knitrout}
\caption{Forest plot for the corticosteroids and COVID-19 mortality meta-analysis \citep{who2020corticosteroids}. Shown are 95\% confidence intervals (CI) and confidence densities for the average effect obtained from weighted and unweighted CD-Edgington methods and the Hartung--Knapp--Sidik--Jonkman method, as well as 95\% prediction intervals (PI) and predictive distributions obtained from the proposed PCD-full approach under different weighting schemes, the Higgins--Thompson--Spiegelhalter method, and the parametric bootstrap approach. Confidence and predictive densities are displayed as scaled kernel density estimates based on Monte Carlo samples.}
\label{fig:weighedexample}
\end{figure}

We revisit the case study on corticosteroids and mortality in hospitalized COVID-19 patients presented in Section~2.6 of the main text. We consider two study-specific weighting schemes for the Edgington-based methods: inverse-standard-error weights ($w_i = 1/\widehat\sigma_i$) and inverse-variance weights ($w_i = 1/\widehat\sigma_i^2$). For prediction, we focus on the PCD-full approach, which showed the best overall performance in the simulation study. For reference, we report the HKSJ estimate and Higgins--Thompson--Spiegelhalter (HTS) and parametric bootstrap prediction intervals. For HKSJ and HTS, heterogeneity is estimated using Paule--Mandel. Web Figure~\ref{fig:weighedexample} displays the resulting 95\% confidence and prediction intervals together with the corresponding confidence and predictive distributions. Additionally, Web Tables~\ref{tab:weighted-ci-example} and \ref{tab:weighted-pi-example} summarize the corresponding point estimates, interval bounds, widths, skewness measures, two-sided $p$-values, and estimated confidence probabilities of a non-negative treatment effect in a future study.

Applying study-specific weights shifts point estimates, confidence and prediction intervals towards negative values, i.e., towards the effect estimates of the larger and more precise studies, most notably the RECOVERY trial. In particular, weighting by inverse-standard-errors yields a point estimate (-0.37) very close to the HKSJ estimate (-0.36), whereas inverse-variance weighting produces an even stronger shift (-0.45).  Correspondingly, the associated two-sided $p$-values for the null hypothesis of a zero-effect decrease from 0.39 in the unweighted analysis to 0.12 under inverse-standard-error weighting and 0.12 under inverse-variance weighting. The same leftward shift is visible for confidence and predictive distributions. The confidence interval becomes slightly narrower under inverse-standard-error weighting, but wider under inverse-variance weighting. Prediction intervals remain wide overall. This indicates that, despite stronger evidence for a beneficial average treatment effect under weighting, substantial uncertainty remains regarding the effect in a future study. Both confidence and prediction intervals remain skewed under the two weighting approaches; however, stronger weighting reduces the degree of skewness, resulting in more symmetric intervals.

\begin{table}[ht]
\centering
\caption{Point estimates and 95\% confidence intervals (CI) for the average treatment effect (log odds ratio) for the corticosteroids and COVID-19 mortality meta-analysis \citep{who2020corticosteroids}. For CD-Edgington, results are reported for unweighted, inverse-standard-error ($w_i = 1/\widehat\sigma_i$), and inverse-variance ($w_i = 1/\widehat\sigma_i^2$) weighting schemes. Results from the Hartung--Knapp--Sidik--Jonkman method are shown for reference. 
      Two-sided $p$-values are reported for testing $H_0: \mu = 0$ against $H_1: \mu \neq 0$.} 
\label{tab:weighted-ci-example}
\begin{tabular}{lrrrrr}
  \toprule
Method & Estimate & 95\% CI & Width & Skewness & p-value \\ 
  \midrule
Hartung--Knapp--Sidik--Jonkman & -0.36 & -0.72  to  -0.00 & 0.72 & 0.000 & 0.048 \\ 
  CD-Edgington (unweighted) & -0.18 & -0.61  to  0.46 & 1.07 & 0.200 & 0.39 \\ 
  CD-Edgington ($w_i = 1/\widehat\sigma_i$) & -0.37 & -0.84  to  0.19 & 1.03 & 0.097 & 0.12 \\ 
  CD-Edgington ($w_i = 1/\widehat\sigma_i^2$) & -0.45 & -1.08  to  0.22 & 1.31 & 0.028 & 0.12 \\ 
  \end{tabular}
\end{table}
\begin{table}[ht]
\centering
\caption{95\% prediction intervals (PI) and medians of predictive distributions for the treatment effect in a future study (log odds ratio) for the corticosteroids and COVID-19 mortality meta-analysis \citep{who2020corticosteroids}. PCD-full results are shown for unweighted, inverse-standard-error ($w_i = 1/\widehat\sigma_i$), and inverse-variance ($w_i = 1/\widehat\sigma_i^2$) weighting schemes. The Higgins--Thompson--Spiegelhalter and parametric bootstrap methods are included for comparison.} 
\label{tab:weighted-pi-example}
\begin{tabular}{lrrrrr}
  \toprule
Method & Median & 95\% PI & Width & Skewness & Conf($\theta_{new} \ge 0$) \\ 
  \midrule
Higgins--Thompson--Spiegelhalter & -0.36 & -0.95  to  0.23 & 1.18 & 0.000 & 0.088 \\ 
  Parametric bootstrap & -0.38 & -1.12  to  0.64 & 1.76 & 0.153 & 0.137 \\ 
  PCD-full (unweighted) & -0.23 & -1.45  to  1.25 & 2.70 & 0.101 & 0.262 \\ 
  PCD-full ($w_i = 1/\widehat\sigma_i$) & -0.40 & -1.67  to  1.03 & 2.70 & 0.064 & 0.154 \\ 
  PCD-full ($w_i = 1/\widehat\sigma_i^2$) & -0.46 & -1.82  to  1.00 & 2.83 & 0.037 & 0.136 \\ 
  \end{tabular}
\end{table}

\end{document}